\documentclass[usenatbib]{mnras}

\usepackage{graphicx}
\usepackage{hyperref}
\usepackage{xspace}
\usepackage{amsmath}
\usepackage{mathtools}

\def\psii{\psi'}
\def\psiii{\psi''}
\def\psiiii{\psi'''}
\def\psiiv{\psi''''}

\def\tein{\theta_{\mathrm{E}}}
\def\reff{R_e}
\def\sigmae{\sigma_{e2}}

\def\gammadm{\gamma_{\mathrm{DM}}}

\def\betamu{\beta_{|\mu|=1}}

\def\Hunit{\,\rm{km}\,{s}^{-1}\,\rm{Mpc}^{-1}}

\def\Sref#1{Section~\ref{#1}\xspace}
\def\Fref#1{Figure~\ref{#1}\xspace}

\def\Eref#1{Equation~\ref{#1}\xspace}

\begin{document}

\title{On the Choice of Lens Density Profile in Time Delay Cosmography}
\author[Sonnenfeld]{
Alessandro~Sonnenfeld$^{1}$\thanks{E-mail:alessandro.sonnenfeld@ipmu.jp}
\\
$^{1}$Kavli IPMU (WPI), UTIAS, The University of Tokyo, Kashiwa, Chiba 277-8583, Japan \\
}

\maketitle

\begin{abstract}
Time delay lensing is a mature and competitive cosmological probe.
However, it is limited in accuracy by the well-known problem of the mass-sheet degeneracy:
too rigid assumptions on the density profile of the lens can potentially bias the inference on cosmological parameters.
I investigate the degeneracy between the choice of the lens density profile and the inference on the Hubble constant, focusing on double image systems.
By expanding lensing observables in terms of the local derivatives of the lens potential around the Einstein radius, and assuming circular symmetry,
I show that three degrees of freedom in the radial direction are necessary to achieve a few percent accuracy in the time-delay distance.
Additionally, while the time delay is strongly dependent on the second derivative of the potential, observables typically used to constrain lens models in time-delay studies, such as image position and radial magnification information, are mostly sensitive to the first and third derivatives, making it very challenging to accurately determine time-delay distances with lensing data alone.
Tests on mock observations show that the assumption of a power-law density profile results in a 5\% average bias on $H_0$, with a 6\% scatter.
Using a more flexible model and adding unbiased velocity dispersion constraints allows to obtain an inference with 1\% accuracy.
A power-law model can still provide 3\% accuracy if velocity dispersion measurements are used to constrain its slope.
Although this study is based on the assumption of axisymmetry, its main findings can be generalised to cases with moderate ellipticity.
\end{abstract}


\section{Introduction}\label{sect:intro}

Time delay cosmography, the measurement of a difference in the travel time between light rays corresponding to multiple images of a strongly lensed variable source, is a powerful cosmological probe. Thanks to the simplicity of its underlying physics, purely gravitational, it can provide estimates of cosmological parameters, such as the expansion rate of the Universe, with relatively few observations \citep[see][for a review]{T+M16}.
Recent years have seen a remarkable improvement in the quality of the data and the analysis techniques used in time delay lensing, thanks to the efforts of the H0LiCOW collaboration \citep{Suy++17}.
The current precision in the determination of $H_0$ from time delay lensing is $3.8\%$, assuming a flat Universe \citep{Bon++17}.
This result was obtained by using only three quadruply imaged quasars, leaving enormous potential for improvement with the discovery and analysis of new lenses.

As precision improves, the requirements on accuracy become increasingly more stringent.
One of the caveats of time delay cosmography is that, in order to make an accurate inference on the cosmological parameters, the lens potential in the proximity of the quasar images, and the position of the quasar in the source plane, need to be accurately estimated.
This is typically done by fitting simple parametrized models to the image positions of the strongly lensed quasar and the surface brightness distribution of its host galaxy.
However, the determination of the properties of the lens is complicated by the mass-sheet degeneracy: given a lens model, we can obtain a continuous family of models that have the same image positions and flux ratios as the original one, but with different time delays \citep{FGS85}.

Typically, power-law density profiles are used in the modelling of time delay lenses.
However, since the mass-sheet transformation of a power-law lens is no longer a power-law, such a choice can artificially restrict the range of viable models to be used in the analysis, leading to an underestimate of the uncertainty.
\citet{S+S13} showed that the assumption of a power-law density profile can in principle introduce a large bias in the inferred value of the Hubble constant, in cases when the true density profile is not a pure power-law.
The amount of the bias on the determination of $H_0$ due to the power-law assumption was further quantified by \citet{Xu++16}, using simulated lenses extracted from a cosmological hydrodynamical simulation.
Here I investigate further this issue, from a different point of view.

The aim of this work is to determine what is the minimum number of degrees of freedom in the lens density profile in order to reproduce lensing observables and time delays within a given accuracy goal.
I limit this study to double image systems. Furthermore, I assume circular symmetry throughout this work. Most of the results derived here, however, can be generalised to cases with moderate departures from axisymmetry.

The structure of this paper is the following. In \Sref{sect:pot}, I derive a Taylor expansion of the lens potential and of the main lensing observables around the Einstein radius of a circular lens with a generic density profile.
In \Sref{sect:pl} I focus on power-law lens models, and use the analytical results derived in \Sref{sect:pot} to point out possible issues with this family of density profiles.
In \Sref{sect:mock} I generate simple mock time delay lensing observations, fit power-law models to them, and quantify the bias in the determination of $H_0$.
In \Sref{sect:gnfw} I attempt to gain in accuracy by increasing the number of degrees of freedom of the model, and by adding velocity dispersion constraints.
I discuss the results in \Sref{sect:discuss} and conclude in \Sref{sect:concl}.
For the generation of mock observations, I assume a flat Lambda CDM cosmology with $\Omega_M = 0.3$ and $H_0 = 70\,\rm{km}\,{s}^{-1}\,\rm{Mpc}^{-1}$.

\section{Lensing observables and lens potential derivatives}\label{sect:pot}

The aim of this Section is to build intuition over what aspects of the lens potential are different lensing observables most sensitive to, in a similar spirit to the work by \citet{Koc02}.
I will consider a circularly symmetric lens, and express various lensing observables in terms of the local derivatives of the potential at the Einstein radius.
The mapping between the lens plane and a source located just outside of the tangential caustic of an axisymmetric lens with a perturber, has been derived by \citet{Wag17b} to leading order in the Taylor series of the lens potential.
Here I adopt a similar formalism to obtain image positions, time delays and radial magnification ratios of doubly imaged sources in the purely axisymmetric case, up to the second nontrivial order in the Taylor expansion of the potential \citep[see also][for a similar study in the case of merging triplets and pairs of images]{W+B16, Wag17}.
%

Let $\beta$ and $\theta$ be angular coordinates in the source and image plane respectively, with origin at the optical axis and same orientation.
Let $\psi(\theta)$ denote the lens potential and $\tein$ be the Einstein radius.
Let us consider a multiply-imaged source at position $\beta_s>0$.
The corresponding image positions are determined by the lens equation:
\begin{equation}\label{eq:lensequation}
\beta_s = \theta  - \alpha(\theta).
\end{equation}
For non-singular lenses, three images will form: one at $\theta_A > \tein$, and the other two at $-\tein < \theta_B < \theta_C < 0$.
The image at $\theta_C$ is, usually, highly de-magnified, and will be ignored from now on.
For small values of $\beta_s$, images A and B will be close to the critical curve. It is then possible to do a Taylor expansion of the lens equation around $\theta=\tein$.
The lens potential at a position $\theta_A = \tein + \Delta\theta_A$ is, to fourth order in $\Delta\theta_A$,
\begin{equation}\label{eq:potA}
\psi(\theta_A) = \psi(\tein) + \psii \Delta\theta_A + \frac12 \psiii \Delta\theta_A^2 + \frac16 \psiiii \Delta\theta_A ^ 3 + \frac{1}{24}\psiiv\Delta\theta_A^4,
\end{equation}
where $\psii$, $\psiii$, etc. are derivatives of the potential evaluated at the Einstein radius:
\begin{align}
\psii &\equiv \left. \frac{d\psi}{d\theta} \right\rvert_{\theta=\tein}, \nonumber \\
\vdotswithin{\psiiv} & \vdotswithin{\left. \frac{d^4\psi}{d\theta^4} \right\rvert_{\theta=\tein}} \\
\psiiv &\equiv \left. \frac{d^4\psi}{d\theta^4} \right\rvert_{\theta=\tein} \nonumber.
\end{align}
The first derivative of the lens potential is the deflection angle $\alpha(\theta)$, which, evaluated at the Einstein radius, is equal to the Einstein radius itself:
\begin{equation}\label{eq:rein}
\psii = \tein.
\end{equation}
Similarly to \Eref{eq:potA}, the potential at a position $\theta_B = -\tein + \Delta\theta_B$ is, to fourth order in $\Delta\theta_B$,
\begin{equation}\label{eq:potB}
\psi(\theta_B) = \psi(\tein) - \psii \Delta\theta_B + \frac12 \psiii \Delta\theta_B^2 - \frac16 \psiiii \Delta\theta_B ^ 3 + \frac{1}{24}\psiiv\Delta\theta_B^4,
\end{equation}
where I have used the symmetry of the lens potential around $\theta = 0$ (i.e. the circular symmetry of the lens).
I will now proceed to obtain an expression relating the two image displacements, $\Delta\theta_A$ and $\Delta\theta_B$.
I will only consider terms up to second order in $\Delta\theta_A$, since, as will be shown, this is sufficient to obtain a description of the time delay with a few percent systematic error.
With this choice, the lens equation for image A reads
\begin{equation}
\beta_s = \theta_A - \psii - \psiii \Delta\theta_A - \frac12 \psiiii \Delta\theta_A^2.
\end{equation}
Using $\theta_A = \tein + \Delta\theta_A$ and \Eref{eq:rein}, this simplifies to
\begin{equation}\label{eq:imA}
\beta_s = \Delta\theta_A - \psiii \Delta\theta_A - \frac12 \psiiii \Delta\theta_A^2.
\end{equation}
For image B, the same procedure leads to
\begin{equation}\label{eq:imB}
\beta_s = \Delta\theta_B - \psiii \Delta\theta_B + \frac12 \psiiii \Delta\theta_B^2.
\end{equation}
By equating the right hand sides of \Eref{eq:imA} and \Eref{eq:imB}, we can obtain the following second order equation in $\Delta\theta_A$ and $\Delta\theta_B$:
\begin{equation}
\frac12\psiiii\Delta\theta_B^2 + (1-\psiii)\Delta\theta_B = -\frac12\psiiii\Delta\theta_A^2 + (1-\psiii)\Delta\theta_A.
\end{equation}
Dividing both sides by $(1-\psiii)$ and solving for $\Delta\theta_B$, one obtains
\begin{align}\label{eq:quadratic}
\Delta\theta_B = & \frac{1-\psiii}{\psiiii}\times \nonumber \\
& \left[-1 \pm \sqrt{1 + 2\frac{\psiiii}{1-\psiii}\Delta\theta_A - \left(\frac{\psiiii}{1-\psiii}\right)^2\Delta\theta_A^2}\right].
\end{align}
Of the two solutions, only the one corresponding to the plus sign is physically meaningful: the other solution does not go to zero in the limit $\Delta\theta_A \rightarrow 0$.
Taylor expansion of the above equation to second order in $\Delta\theta_A$ gives
\begin{equation}\label{eq:lens2nd}
\Delta\theta_B = \Delta\theta_A - \frac{\psiiii}{1-\psiii}\Delta\theta_A^2 + O(\max{\{\Delta\theta_A^3, \Delta\theta_B^3\}}).
\end{equation}
A full derivation of the equation above is provided in Appendix~\ref{sect:appendixa}.
By bringing $(1-\psiii)$ to the denominator, I have assumed that $\psiii\neq1$.
This is typically the case for observed galaxy-scale lenses, for which the total density profile is observed to be close to isothermal, corresponding to $\psiii\approx0$ (see also \Fref{fig:psi}).

\Eref{eq:lens2nd} relates the position of image B to that of image A and the derivatives of the potential.
The dependence on the Einstein radius is implicit in the definition of $\Delta\theta_A$ and $\Delta\theta_B$.
Let us now examine how different {\em observables} depend on the lens potential derivatives.
Let us start with the image separation. To second order in $\Delta\theta_A$, this reads
\begin{equation}\label{eq:imsep}
\theta_A - \theta_B = 2\psii + \frac{\psiiii}{1-\psiii}\Delta\theta_A^2,
\end{equation}
that is, the image separation is approximately equal to twice the Einstein radius.

Let us now consider the magnification. Absolute magnification is only observable if the lensed source is a standard candle, or a standard ruler.
The ratio of magnifications of different images of the same source is more easily accessible. Still, in the case of lensed quasars, magnification ratios are difficult to interpret in the context of smooth mass models, due to microlensing or the presence of substructure. 
Nevertheless, some constraints on magnification can be obtained by careful modelling of the images of a lensed extended source.
In detailed time delay lens studies, for instance, the slope of the density profile of the lens is usually constrained by fitting a parametrized model to the full surface brightness distribution of the quasar host galaxy \citep[see e.g.][]{Suy12}. 
This constraint from extended source modelling is essentially a measurement of radial magnification ratio. 
When an extended source is lensed into two arcs at two different distances from the lens, the ratio between the arc widths, which is directly observable, is given by the ratio of radial magnifications at the two positions.

Radial magnification, for a circular lens, is defined as
\begin{equation}\label{eq:murdef}
\mu_r = \left(\frac{d\beta}{d\theta}\right)^{-1} = \left(1 - \frac{d\alpha}{d\theta}\right)^{-1}.
\end{equation}
The Taylor expansion around image A of the derivative of the deflection angle is,
\begin{equation}\label{eq:dalphaa}
\frac{d\alpha(\theta_A)}{d\theta} = \psiii + \psiiii\Delta\theta_A + \psiiv\Delta\theta_A^2 + O(\Delta\theta_A^3).
\end{equation}
For image B,
\begin{equation}\label{eq:dalphab}
\frac{d\alpha(\theta_B)}{d\theta} = \psiii - \psiiii\Delta\theta_B + \psiiv\Delta\theta_B^2 + O(\Delta\theta_B^3).
\end{equation}
Radial magnification introduces the fourth order derivative of the potential. This, however, cancels out when taking the ratio between the magnifications at image A and B. The result is the following:
\begin{equation}\label{eq:radmagrat}
\frac{\mu_{r,A}}{\mu_{r,B}} = 1 + \frac{2\psiiii}{1-\psiii}\Delta\theta_A + \left(\frac{\psiiii}{1-\psiii}\right)^2\Delta\theta_A^2 + O(\Delta\theta^3).
\end{equation}
The full derivation is shown in Appendix~\ref{sect:appendixb}.

Finally, let us consider the time delay between image B and A. This is given by
\begin{equation}\label{eq:dt}
\Delta t = \frac{D_{\Delta t}}{c}\left[\frac{(\theta_B - \beta_s)^2}{2} - \psi(\theta_B) - \frac{(\theta_A - \beta_s)^2}{2} + \psi(\theta_A)\right],
\end{equation}
where $c$ is the speed of light, and $D_{\Delta t}$ is the so-called time-delay distance. This, in turn, is defined as
\begin{equation}
D_{\Delta t} \equiv (1+z_d) \frac{D_d D_s}{D_{ds}},
\end{equation}
where $z_d$ is the redshift of the lens, $D_d$, $D_s$ are the angular diameter distances of the lens and source relative to the observer, and $D_{ds}$ the angular diameter distance between lens and source. 

Substituting Equations \ref{eq:potA}, \ref{eq:potB}, \ref{eq:imA}, \ref{eq:imB} and \ref{eq:lens2nd} into \Eref{eq:dt}, and keeping terms up to second order in $\Delta\theta_A$, gives
\begin{equation}\label{eq:dt2nd}
\Delta t = 2\frac{D_{\Delta t}}{c}\psii(1 - \psiii)\Delta\theta_A\left[1 + \frac{\psiiii}{2(1 - \psiii)}\Delta\theta_A\right].
\end{equation}
To leading order in image displacement, the time delay depends linearly on both the first and second derivatives of the lens potential. The third order derivative only enters \Eref{eq:dt2nd} at the second order in $\Delta\theta_A$. A step-by-step derivation is provided in Appendix~\ref{sect:appendixc}.

Let us now use \Eref{eq:dt2nd} to make quantitative statements on how uncertainties on the different derivatives of the potential translate into uncertainties in the time-delay.
Let us consider, for example, a fiducial image configuration such as $\Delta\theta_A = 0.2\tein$.
Assuming, for simplicity, that the Einstein radius is known exactly \citep[this is a good approximation, since $\tein$ is typically determined with better than 1\% precision in time-delay lens studies][]{Suy++14, Won++17}, an error on the second derivative propagates linearly into an error on the time-delay. Assuming the second derivative is also known exactly, and assuming that the product $\psii\psiiii$ is of order unity, errors on $\psiiii$ can introduce errors on the time delay on the order of 20\%.
This value should be compared with the few percent precision attainable today in time delay cosmography studies \citep{Suy++17}.

\Eref{eq:dt2nd} itself is accurate to 4\% for the fiducial image configuration, assuming the coefficient of the $\Delta\theta_A^3$ term in the Taylor series to be on the order of $D_{\Delta t}/(c\tein^3)$.
We can then conclude that, in order to make a time delay cosmology measurement with an accuracy of a few percent, a lens model density profile with at least three degrees of freedom is required.

\subsection{Impact of the mass-sheet degeneracy}\label{ssec:msd}

As can be seen from \Eref{eq:imsep} and \Eref{eq:radmagrat}, 
the observables typically used to constrain lens models, image separation and radial magnification ratio, appear to depend on a fixed combination of the second and third derivative of the potential, $\psiiii/(1-\psiii)$.
This suggests that, for a given lens potential $\psi(\theta)$, a different potential that keeps the ratio $\psiiii/(1-\psiii)$ invariant will produce the same observables.
This invariance property is the mass-sheet degeneracy. Since the lens potential is the Laplacian of the surface mass density,
\begin{equation}
\Delta\psi = 2\kappa,
\end{equation}
a transformation of the kind $\kappa \rightarrow \lambda\kappa + (1-\lambda)$ corresponds to the following transformation in the potential:
\begin{equation}\label{eq:pottransform}
\phi(\theta) = \lambda\psi(\theta) + \frac{(1-\lambda)}{2}\theta^2.
\end{equation}
The new potential $\phi$ satisfies the following equality
\begin{equation}\label{eq:mymst}
\frac{\phi'''}{1-\phi''} = \frac{\psiiii}{1-\psiii},
\end{equation}
which means that it will predict the same radial magnifications and image positions as the old potential, for a corresponding change in source position. Such a change in source position will modify the predicted time delay.

\Eref{eq:mymst} is also satisfied by the more general source-position transformation \citep[SPT][]{S+S14}, since the SPT is equivalent to the mass-sheet transformation when derivatives of the potential up to third order are considered \citep[see subsection 3.2 of][]{S+S14}.

\subsection{The moderate ellipticity case}

The equations derived so far are strictly true only under the assumption of circular symmetry.
Axisymmetric lenses are qualitatively different from more realistic models.
For instance, their critical curves, isopotential and isodensity contours have all the same shape.
One consequence of the departure from axisymmetry is that, for doubles, as the source crosses the tangential caustic, only one of the two images touches the critical curve.
In other words, as $\Delta\theta_A$, interpreted as distance from the tangential critical curve in an appropriate coordinate frame, goes to zero, $\Delta\theta_B$ approaches a finite value, or vice versa, depending on the orientation of the source position with respect to the major axis of the lens.
Nevertheless, the main results, i.e. the dependence of different observables on different radial derivatives of the lens potential, are still qualitatively correct for lenses with moderate ellipticity.

The time delay is still proportional, to leading order in image displacement from the critical curve, to the difference in deflection angles at the location of the two images.
This introduces a dependence on the second derivative of the potential.
The radial magnification is, to first order in image position, given by the value of the radial eigenvalue of the lens Jacobian matrix on the critical curve, plus the scalar product between the image displacement vector and the gradient of the radial eigenvalue.
This gradient introduces a dependence on the third derivative of the potential, which does not cancel out when taking the ratio between $\mu_{r,A}$ and $\mu_{r,B}$.

Let us verify this conjecture explicitly for a simple case, to lowest order in image displacement from the tangential critical curve.
Let us consider a lens with an elliptically symmetric potential:
\begin{equation}\label{eq:potell}
\Phi(\theta_1, \theta_2) = \phi(\rho),
\end{equation} 
with
\begin{equation}
\rho^2 \equiv q\theta_1^2 + \frac1q\theta_2^2,
\end{equation}
where $q$ is the axis ratio of the isopotential curves.
Let us consider, for simplicity, a source aligned with the major axis of the lens:
\begin{equation}
\boldsymbol\beta = (\beta_s, 0),
\end{equation}
with $\beta_s > 0$.
Let $\boldsymbol\beta$ be outside of the tangential caustic, and let $\theta_A$ and $\theta_B$ be the coordinates of the two images in the $\theta_1$ axis, with $\theta_A > 0$ and $\theta_B < 0$.
Finally, let $\Delta\theta_A$ and $\Delta\theta_B$ be the difference between the $\theta_1$ coordinates of the two images and the points $(\theta_c,0)$ and $(-\theta_c,0)$ where the tangential critical curve intercepts the $\theta_1$ axis:
\begin{equation}
\Delta\theta_A = \theta_A - \theta_c,
\end{equation}
\begin{equation}
\Delta\theta_B = \theta_B + \theta_c.
\end{equation}
For an elliptical potential, $\theta_c$ is the solution to the following equation
\begin{equation}\label{eq:critell}
\left.\frac{\partial\phi}{\partial\rho}\right\rvert_{\rho=q^{1/2}\theta_c} = q^{3/2}\theta_c,
\end{equation}
derived in Appendix~\ref{sect:appendixe}.
Let us now expand the lens potential in a Taylor series for small displacements in the $\theta_1$ direction around $(\theta_c, 0)$ (image A) and $(-\theta_c, 0)$ (image B).
For image A,
\begin{align}
\Phi(\theta_A, 0) = &\Phi(\theta_c, 0) + \left.\frac{\partial\Phi}{\partial\theta_1}\right\rvert_{(\theta_c,0)}\Delta\theta_A + \frac12\left.\frac{\partial^2\Phi}{\partial\theta_1^2}\right\rvert_{(\theta_c,0)}\Delta\theta_A^2 + \nonumber \\
& + \frac16\left.\frac{\partial^3\Phi}{\partial\theta_1^3}\right\rvert_{(\theta_c,0)}\Delta\theta_A^3 + O(\Delta\theta_A^4).
\end{align}
For image B,
\begin{align}
\Phi(\theta_B, 0) = &\Phi(\theta_c, 0) - \left.\frac{\partial\Phi}{\partial\theta_1}\right\rvert_{(\theta_c,0)}\Delta\theta_B + \frac12\left.\frac{\partial^2\Phi}{\partial\theta_1^2}\right\rvert_{(\theta_c,0)}\Delta\theta_B^2 + \nonumber \\
& - \frac16\left.\frac{\partial^3\Phi}{\partial\theta_1^3}\right\rvert_{(\theta_c,0)}\Delta\theta_B^3 + O(\Delta\theta_B^4),
\end{align}
where I have used the symmetry of $\Phi(\theta_1,0)$ around the origin.
The three partial derivatives, evaluated at $(\theta_c,0)$, are
\begin{align}
\left.\frac{\partial\Phi}{\partial\theta_1}\right\rvert_{(\theta_c,0)} = & q^{1/2}\left.\frac{\partial\phi}{\partial\rho}\right\rvert_{\rho=q^{1/2}\theta_c} \equiv q^{1/2}\phi' \nonumber \\
\left.\frac{\partial^2\Phi}{\partial\theta_1^2}\right\rvert_{(\theta_c,0)} = & q\left.\frac{\partial^2\phi}{\partial\rho^2}\right\rvert_{\rho=q^{1/2}\theta_c} \equiv q\phi'' \\
\left.\frac{\partial^3\Phi}{\partial\theta_1^3}\right\rvert_{(\theta_c,0)} = & q^{3/2}\left.\frac{\partial^3\phi}{\partial\rho^3}\right\rvert_{\rho=q^{1/2}\theta_c} \equiv q^{3/2}\phi''' \nonumber .
\end{align}
The lens equation in the general case is
\begin{equation}\label{eq:2dlens}
\boldsymbol\beta = \boldsymbol\theta - \boldsymbol\alpha(\boldsymbol\theta) = \boldsymbol\theta - \boldsymbol\nabla\Phi.
\end{equation}
Equating the right hand sides of the $\theta_1$ component of \Eref{eq:2dlens} for image A and B, and keeping only terms up to first order in $\Delta\theta_A$ and $\Delta\theta_B$ gives
\begin{equation}
(1-q\phi'')\Delta\theta_B = 2\theta_C - 2q^{1/2}\phi' + \Delta\theta_A(1 - q\phi'').
\end{equation}
As anticipated, $\Delta\theta_B$ no longer goes to zero as $\Delta\theta_A$ goes to zero, but converges to the quantity 
\begin{equation}\label{eq:epsilonq}
2\epsilon_q \equiv 2\frac{\theta_c - q^{1/2}\phi'}{1-q\phi''},
\end{equation}
which vanishes in the limit $q\rightarrow 1$.
Assuming that $\epsilon_q$ is an infinitesimal smaller than $\Delta\theta_A^2$, a similar calculation to the one that led to \Eref{eq:dt2nd}, but limited to the first order in $\Delta\theta_A$, gives the following expression for the time delay between the two images:
\begin{equation}\label{eq:ellpot_dt}
\Delta t = 2\frac{D_{\Delta t}}{c}q^{1/2}\phi'(1 - q\phi'')(\Delta\theta_A + \epsilon_q) + O(\Delta\theta_A^2).
\end{equation}
Also in this case, the time delay depends on the second derivative of the potential \citep[see also][]{Koc02}.
Finally, the radial magnification ratio is given by
\begin{equation}\label{eq:ellpot_radmagrat}
\frac{\mu_{r,A}}{\mu_{r,B}} \approx 1 + 2q^{3/2}\frac{\phi'''}{1-q\phi''}(\Delta\theta_A + \epsilon_q),
\end{equation}
which, to lowest order in $\Delta\theta_A$ and for small values of $\epsilon_q$ is sensitive to the quantity $\phi'''/(1-q\phi'')$.
Equivalent expressions for the case in which the source lies in correspondence to the minor axis of the lens can be obtained by substituting $q \rightarrow 1/q$.
Providing expressions for the general case in which the source is located at an arbitrary position outside the tangential caustic is beyond the goal of this work.

\begin{figure*}
 \begin{tabular}{cc}
 \includegraphics[width=\columnwidth]{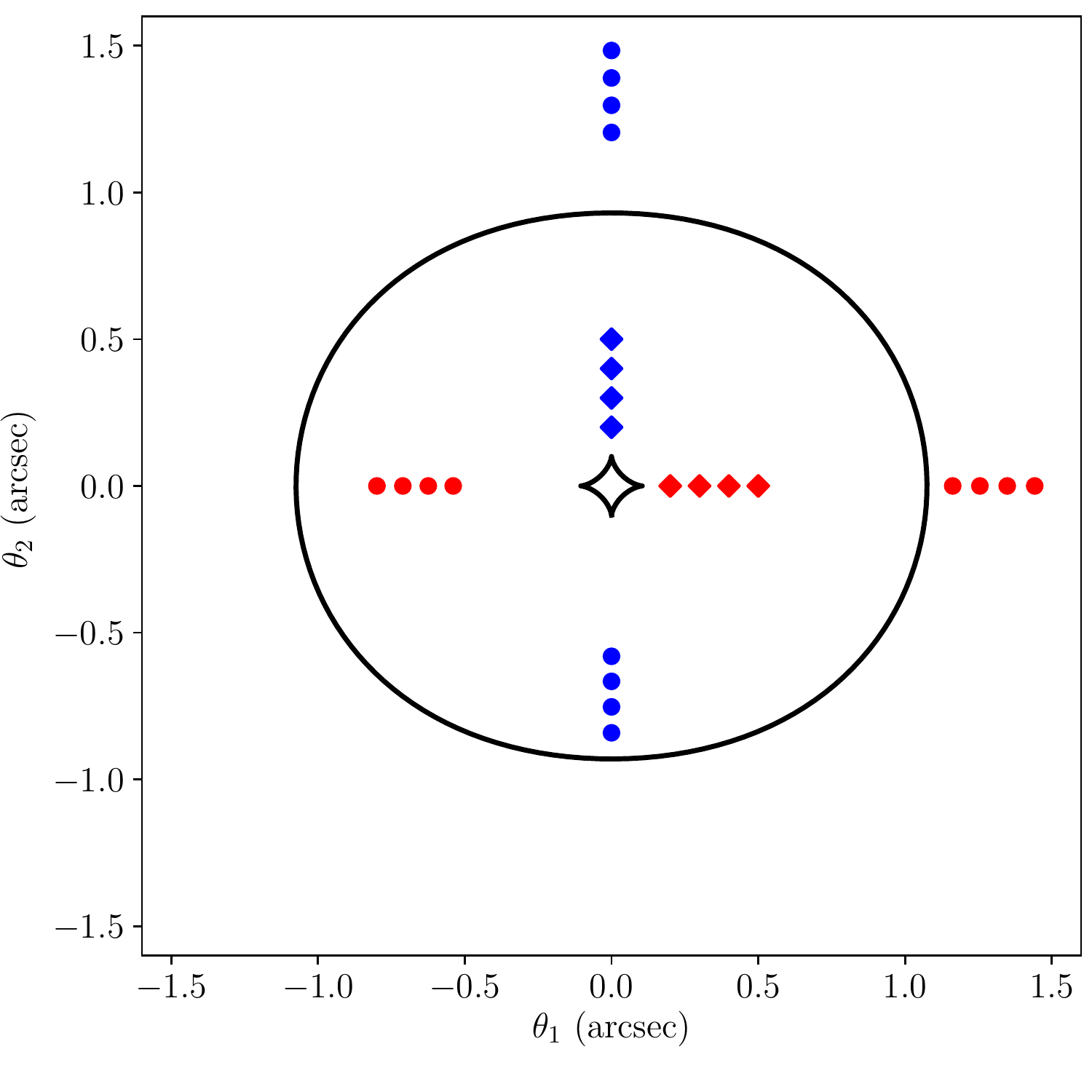} &
 \includegraphics[width=\columnwidth]{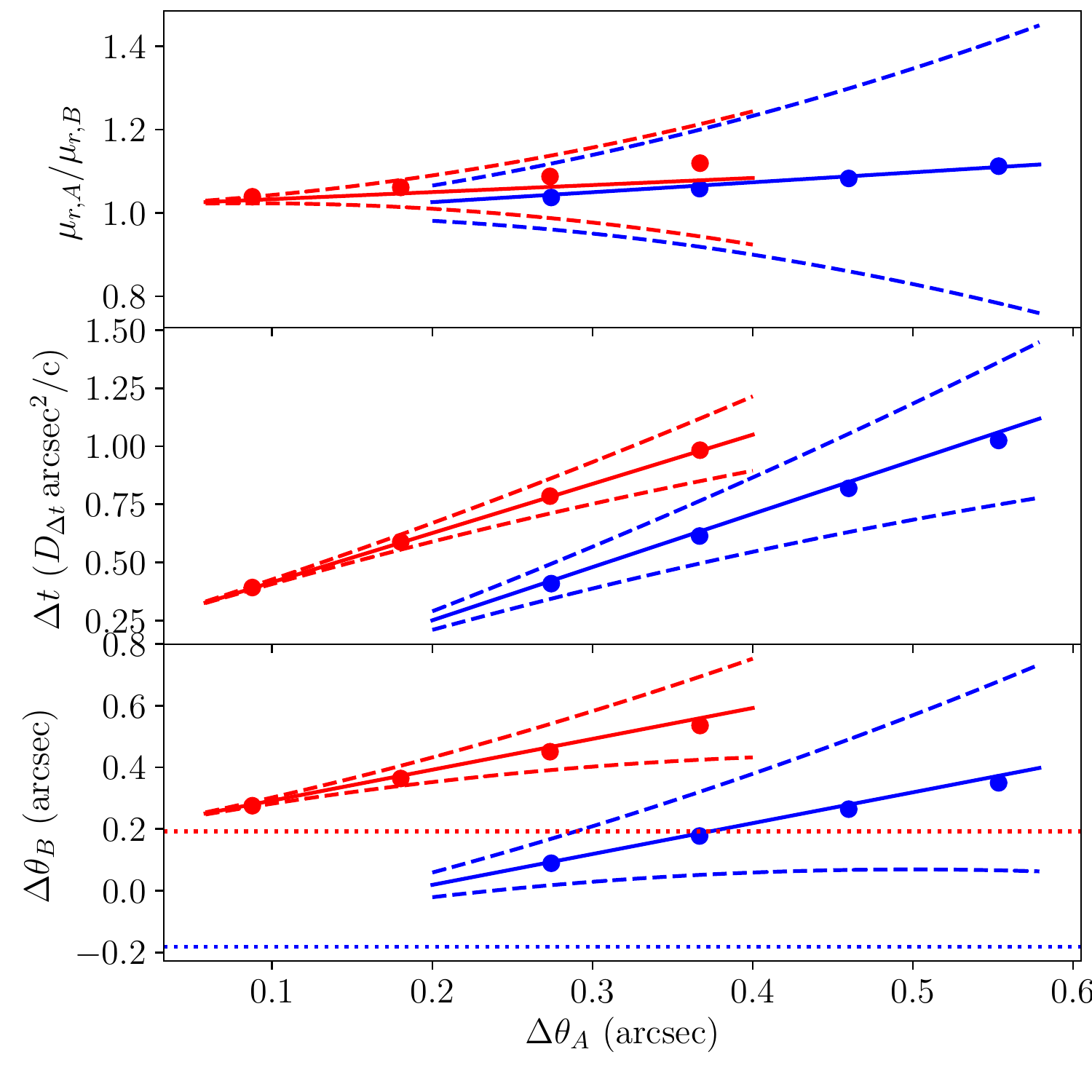}
 \end{tabular}
 \caption{Lens with elliptical potential, with $q=0.95$ and steeper than isothermal radial profile, as specified by \Eref{eq:ellpot_powerlawpot}. {\em Left panel. Solid lines:} tangential critical curve and caustic. {\em Diamonds:} source positions. {\em Circles:} image positions corresponding to each source position.
{\em Right panel:} displacement of image B with respect to the tangential critical curve (bottom row), time delay (middle row), and radial magnification ratio (top row) for the image pair corresponding to each source, as a function of $\Delta\theta_A$.
Red (blue) coloured points correspond to sources placed along the major (minor) axis of the potential. 
Analytical estimates of $\Delta\theta_B$, $\Delta t$ and $\mu_{r,A}/\mu_{r,B}$, obtained by Taylor expansion to first order in $\Delta\theta_A$, are plotted as solid lines. Dashed lines delimit the expected error due to the truncation of the Taylor series to the first order in $\Delta\theta_A$.
The red dotted line in the bottom panel marks the value $2\epsilon_q$, defined in \Eref{eq:epsilonq}, i.e. the value $\Delta\theta_B$ converges to in the limit $\Delta\theta_A\rightarrow0$ for a source aligned with the major axis of the potential. The blue dotted line indicates the value of $-\Delta\theta_A$ corresponding to $\Delta\theta_B=0$, for a source aligned with the minor axis of the potential, and is obtained by replacing $q$ with $q^{-1}$ in \Eref{eq:epsilonq}.
}
 \label{fig:ellpot}
\end{figure*}

The main results derived in this subsection are verified for an example case, shown in \Fref{fig:ellpot}. I consider a lens with elliptical potential, with $q=0.95$, and sources located at different positions along the major and minor axis of the potential distribution.
The radial component of the potential, $\phi(\rho)$, is chosen to be a power-law
\begin{equation}\label{eq:ellpot_powerlawpot}
\phi(\rho) = \frac{1}{3-\gamma}\rho^{3-\gamma}\,(\rm{arcsec}^2),
\end{equation}
with $\gamma=2.1$, corresponding to a density profile slightly steeper than isothermal.
The tangential critical curve and the corresponding caustic are shown in the left panel of \Fref{fig:ellpot}. In the same panel, source positions are marked with diamonds, and the corresponding image positions are shown as circles of the same colour.

The right panel of \Fref{fig:ellpot} shows the distance from the critical curve of image B, the time delay, and the radial magnification ratio of the two images, for each source position, as a function of $\Delta\theta_A$.
In each row, I also plot the analytical expression for the Taylor expansion of the corresponding quantity, to first order in $\Delta\theta_A$, derived using \Eref{eq:2dlens}, \Eref{eq:ellpot_dt} and \Eref{eq:ellpot_radmagrat}.
For this example case, the true values of the quantities $\Delta\theta_B$, $\Delta t$ and $\mu_{r,A}/\mu_{r,B}$ are well reproduced by the corresponding analytical approximations (solid lines) within their nominal accuracy, $\Delta\theta_A^2$ (dashed lines).

\section{Limitations of power-law models}\label{sect:pl}

Let us go back to axisymmetric lenses.
In the previous Section I have shown how image position, radial magnification ratio and time delay depend on local derivatives of the lens potential around the Einstein radius, for small displacements around a symmetric image configuration.
Here I investigate what happens when a power-law density profile is used to model all three of these observables simultaneously.

The lens potential of a singular power-law density profile with Einstein radius $\tein$ and 3D density slope $\gamma$ is given by
\begin{equation}
\psi_{\mathrm{PL}}(\theta) = \frac{\tein^2}{3-\gamma}\left(\frac{\theta}{\tein}\right)^{3-\gamma}.
\end{equation}
By definition, the first derivative of the potential at $\tein$ is equal to the Einstein radius: 
\begin{equation}
\psi'_{\mathrm{PL}} = \tein.
\end{equation}
The second and third derivatives are given by
\begin{equation}\label{eq:psiiipl}
\psi''_{\mathrm{PL}} = 2 - \gamma,
\end{equation}
and
\begin{equation}
\psi'''_{\mathrm{PL}} = \frac{(2-\gamma)(1-\gamma)}{\tein}.
\end{equation}
A power-law density profile is described fully by only two parameters: $\tein$ and $\gamma$. Therefore, only two of its derivatives are independent. For instance, the third derivative can be written in terms of the first two:
\begin{equation}
\psi'''_{\mathrm{PL}} = \frac{\psi''_{\mathrm{PL}}}{\psi'_{\mathrm{PL}}}(\psi''_{\mathrm{PL}} - 1).
\end{equation}
The above second order equation can be inverted as follows:
\begin{equation}\label{eq:psiiipl_given_psiiii}
\psi''_{\mathrm{PL}} = \frac{1 \pm \sqrt{1 + 4\psi'_{\mathrm{PL}}\psiiii_{\mathrm{PL}}}}{2},
\end{equation}
where the positive solution corresponds to values of the power-law slope $\gamma < 1.5$, and the negative solution to values $\gamma > 1.5$.

In time delay cosmography studies, power-law lens models are typically used \citep{Suy++10, Suy++13, Won++17}.
A power-law lens model is fitted to a high resolution image of a lensed quasar and its host galaxy. By reconstructing the multiple, extended, images of the host, the power-law index of the model can be constrained.
The fitted model is then used to predict the time delay between the quasar images, as a function of cosmological parameters, through \Eref{eq:dt}. Finally, the cosmological parameters are constrained by comparing the predicted time delay with the measured value.

Such a procedure can introduce a bias. As shown in \Sref{sect:pot}, in order to make an accurate prediction of the time delay, an equally accurate knowledge of the second derivative of the potential is required. However, the observables used to constrain the model, image positions and radial magnification ratio, are only sensitive to a fixed combination of the second and third derivative, $\psiiii/(1-\psiii)$, as pointed out in subsection~\ref{ssec:msd}. 
Even worse, for $\psiiii$ close to zero, such data gives no information at all on the second derivative.
The time delay prediction obtained with a power-law model is then mostly the result of a particular assumption on the relation between the second and third derivative of the potential.

Qualitatively, the possibility of introducing a bias by the assumption of a power-law density profile was well established \citep{S+S13}.
This treatment can be seen as an alternative formulation of the mass-sheet degeneracy problem, obtained in terms of the derivatives of the lens potential and how these enter different lensing observables.
In the next section, I try to quantify the importance of this effect by simulating time delay cosmography measurements over a mock population of doubly imaged quasars.

\section{Tests on mock lenses}\label{sect:mock}

I generate 100 mock lenses as follows. I draw halo masses, defined as the mass within a shell enclosing an average mass equal to 200 times the critical density of the Universe, from a Gaussian distribution in the logarithm of the halo mass, $\log{M_h}$, centred at $\mu_h=13.2$ and with dispersion $\sigma_h=0.3$. I then draw stellar masses from a Gaussian distribution in $\log{M_*}$, with a mean that depends on halo mass as
\begin{equation}
\mu_* = 11.4 + 0.6(\log{M_h} - \mu_h),
\end{equation}
and dispersion $\sigma_*=0.1$.
This distribution in halo and stellar masses roughly matches that of existing samples of lenses, as well as current constraints on the stellar-to-halo mass relation \citep{Gav++07, Aug++10, Beh++13}.

I assign a redshift to each lens, drawn from a uniform distribution in the interval $0.2 < z_d < 0.4$. I draw source redshifts from a Gaussian distribution with mean $1.5$ and dispersion $0.5$, truncated between $z_{s,min}=0.7$ and $z_{s,max}=4.0$.
I draw effective radii from a Gaussian mass-size relation, with parameters measured by \citet{New++12} on SDSS galaxies, assuming a Salpeter IMF.

I model each lens as the sum of a circular de Vaucouleurs profile \citep{deV48}, for the stars, and a spherically symmetric Navarro Frenk \& White profile \citep[NFW][]{NFW97}, for the dark matter halo.
I draw the concentration parameters of the dark matter halos assuming a Gaussian mass-concentration relation with parameters from \citet{Mac++08}.

Finally, I assign source positions.
I wish to generate lenses with a realistic image configuration for a time delay cosmography measurement: in particular, I wish to avoid extremely asymmetric configurations, in which the counter-image is very close to the centre and de-magnified.
Therefore, for each lens, I draw the source position from a uniform distribution in $\beta^2$ between 0 and $\betamu$, where $\betamu$ is the smallest value of $\beta$ for which the magnification of image B satisfies $|\mu|=1$. 

\begin{figure}
 \includegraphics[width=\columnwidth]{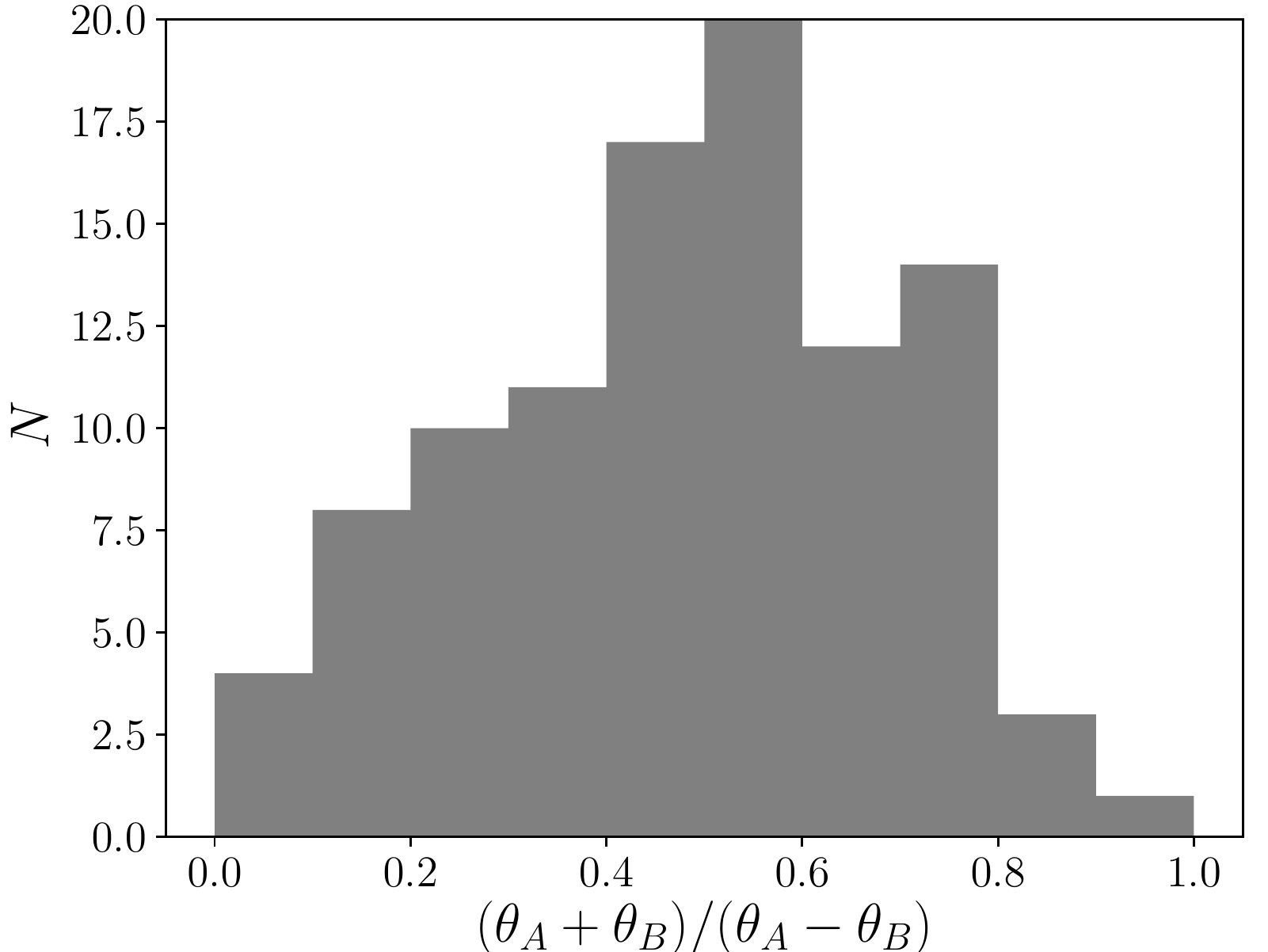}
 \caption{Distribution in the asymmetry of the image configuration for the lenses in the mock sample.}
 \label{fig:asymm}
\end{figure}

In other words, all images of all the lenses in the mock are magnified.
For some lenses, the value of $\betamu$ is not well defined: these are lenses for which the amplitude of the magnification of image B is always larger than unity, for any value of $\beta$ between 0 and the radial caustic. In these cases, I set the radial caustic as the largest allowed value of $\beta$.
\Fref{fig:asymm} plots the distribution in the asymmetry of the image configuration, defined as $(\theta_A + \theta_B)/(\theta_A - \theta_B)$.
This distribution goes well beyond the regime of small displacements around $\tein$, in which the equations derived in \Sref{sect:pot} are guaranteed to be accurate.
Nevertheless, the analysis carried out so far can still provide a valid description of this simulation, as will be shown.

\begin{figure}
 \includegraphics[width=\columnwidth]{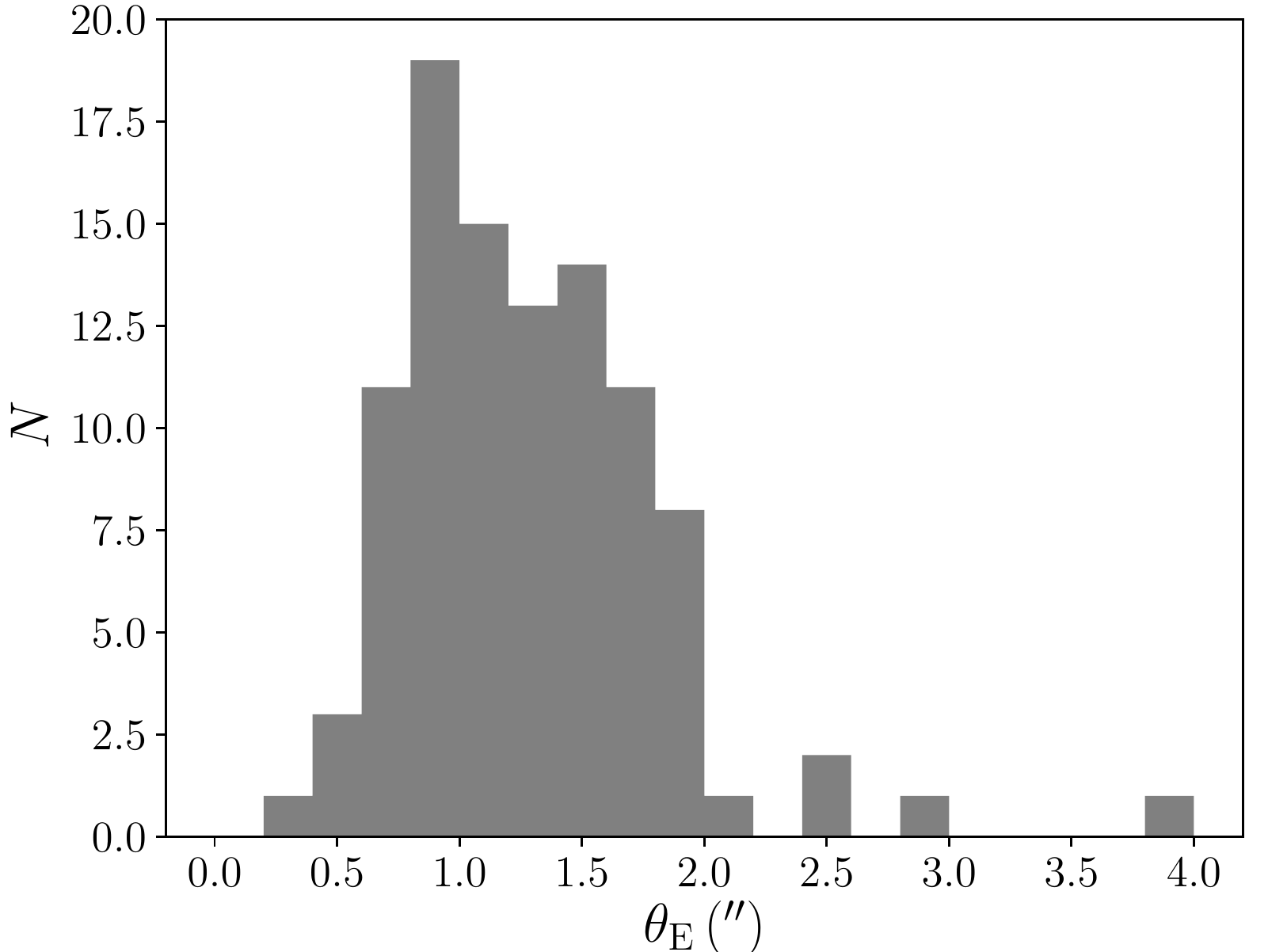}
 \caption{Distribution in Einstein radius, in angular units, for the lenses in the mock sample.}
 \label{fig:rein}
\end{figure}

\begin{figure}
 \includegraphics[width=\columnwidth]{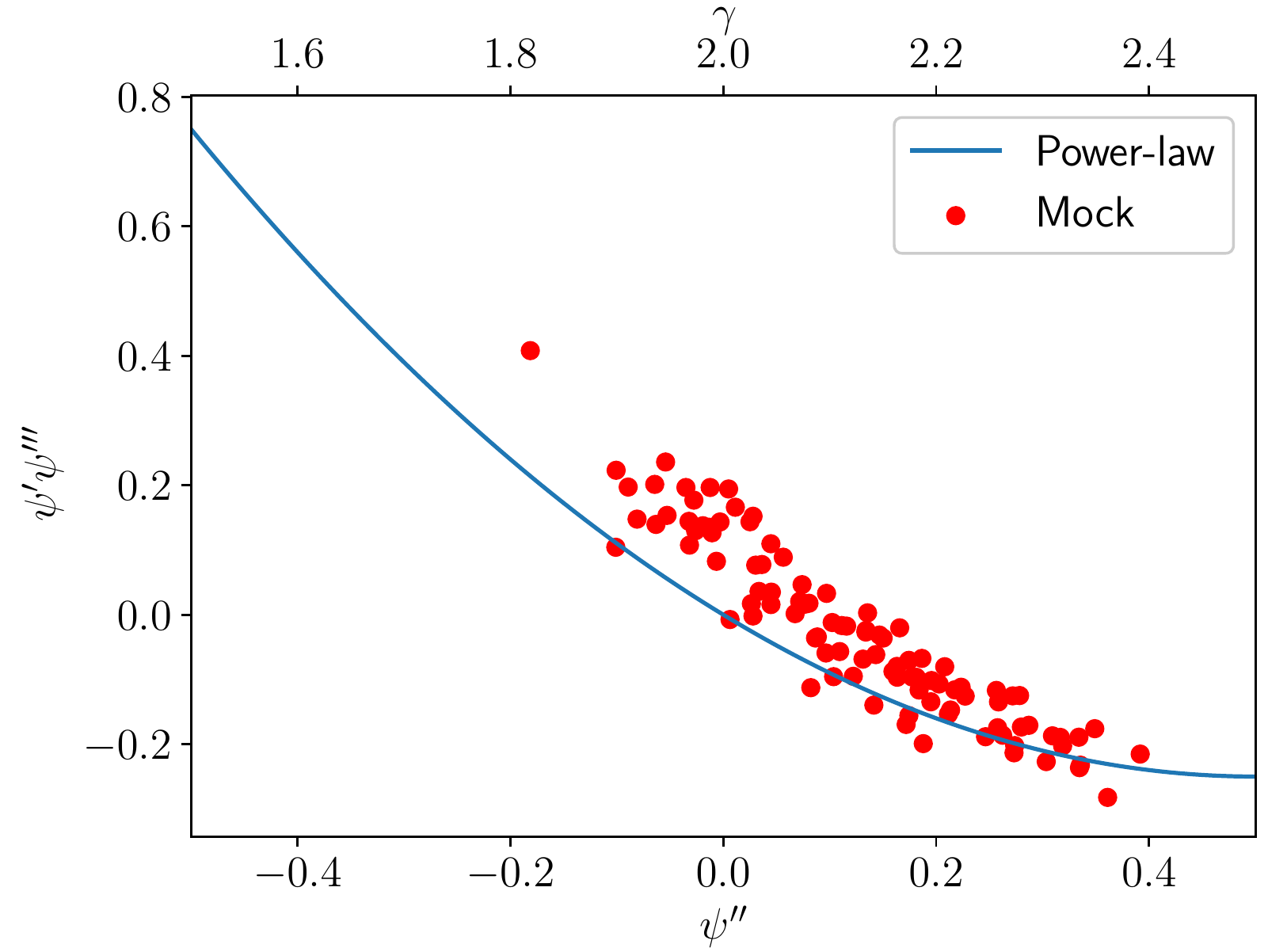}
 \caption{Distribution in $\psiii$ and $\psii\psiiii$, respectively the second derivative of the lens potential at the Einstein radius, and the product between the first and third derivative, for the lenses in the mock sample.
The solid curve shows the relation between $\psiii$ and $\psii\psiiii$ for a power-law density profile. The top x-axis indicates the value of the power-law slope corresponding to a given value of the second derivative.
}
 \label{fig:psi}
\end{figure}

\Fref{fig:rein} plots the distribution in Einstein radius of the sample, while \Fref{fig:psi} shows the second and third derivative of the lens potential at $\tein$, for each lens.
The distribution of lens potential derivatives of the mock galaxies is different from that of a strict power-law profile, though roughly following it.
The farther a lens is from the power-law curve in \Fref{fig:psi}, the larger we expect the bias on the time delay cosmology inference introduced by a power-law assumption to be, with the exact value depending on the image configuration of the system.

\subsection{Power-law model fits}

The goal of this analysis is to quantify the accuracy (or lack thereof) in the determination of cosmological parameters with the assumption of a power-law density profile.
For simplicity, I focus on the Hubble constant, $H_0$, leaving all other cosmological parameters fixed.

For each lens, I fit a power-law density profile to its image positions and image magnification ratios. This step is intended to provide similar constraints on the lens density profile as those obtained by detailed modelling of the surface brightness profile of the quasar host galaxy in real lenses.
I wish to determine what value of $H_0$ one would obtain in the limit of perfect measurements.
Therefore, I assume that image positions, radial magnification ratios and time delays are known exactly.
Under this assumption, the three free parameters of the power-law model, the Einstein radius, the density slope $\gamma$, and the source position, can be determined exactly from the two image positions and radial magnification ratio. 

\begin{figure}
 \includegraphics[width=\columnwidth]{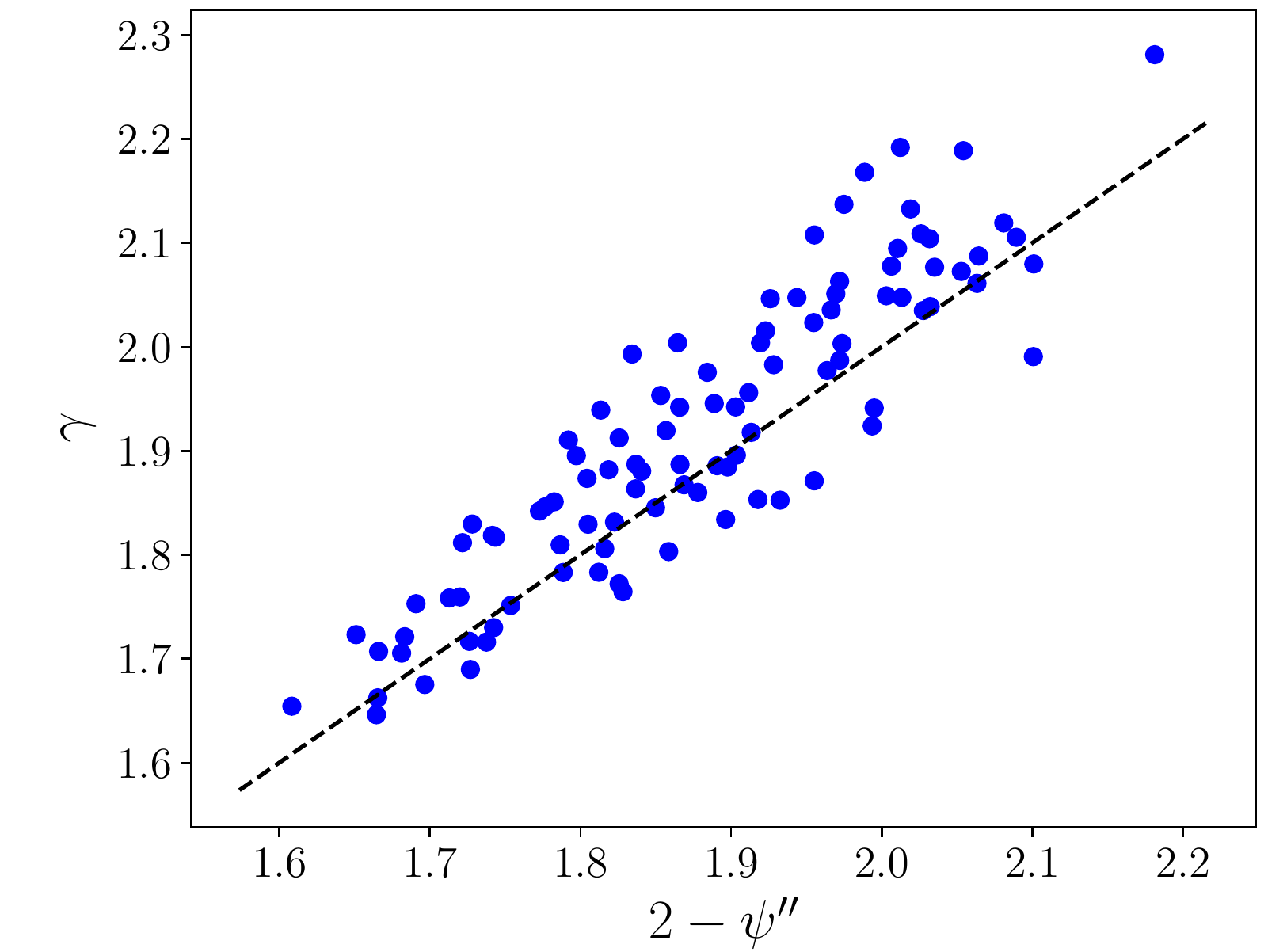}
 \caption{
Inferred values of the density slope $\gamma$, obtained by fitting a power-law density profile to image positions and radial magnification ratios of each mock lens. The values of $\gamma$ are plotted as a function of $2-\psiii$. For a pure power-law density profile we have $\gamma = 2-\psiii$: deviations from the equality line indicate deviations from a pure power-law profile.
}
 \label{fig:slope}
\end{figure}

The values of the inferred density slope are plotted in \Fref{fig:slope}, as a function of $2-\psiii$, where $\psiii$ is the true value of the second derivative of the lens potential at the Einstein radius. For a pure power-law profile, $\gamma = 2-\psiii$ (see \Eref{eq:psiiipl}). Therefore, deviations from the 1-1 line in \Fref{fig:slope} indicate deviations from a pure power-law.

\begin{figure}
 \includegraphics[width=\columnwidth]{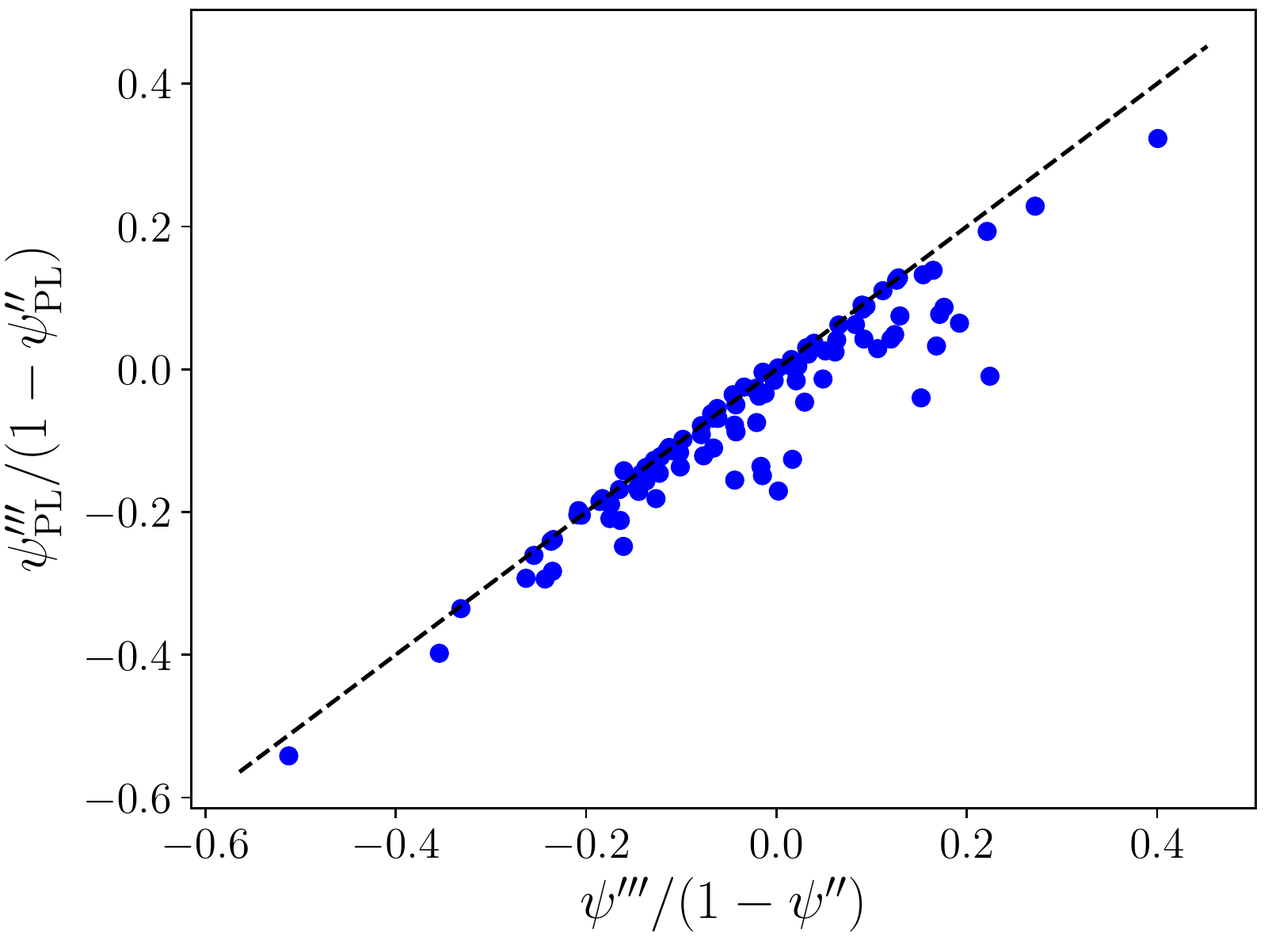}
 \caption{
Inferred value of the combination of derivatives of the lens potential $\psiiii/(1-\psiii)$, obtained by fitting power-law models to image position and radial magnification ratio data, as a function of the true value. 
}
 \label{fig:psicomb}
\end{figure}

The scatter around the equality line in \Fref{fig:slope} indicates, as argued in \Sref{sect:pl}, that power-law lenses cannot provide an accurate estimate of the second derivative of the potential, when fitted to image position and radial magnification ratio data.
Image position and radial magnification observations provide information on the following combination of derivatives, instead:
\begin{equation}\label{eq:psicomb}
\frac{\psiiii}{1 - \psiii},
\end{equation}
as can be seen by examining \Eref{eq:imsep} and \Eref{eq:radmagrat}.
We can expect the quantity $\psiiii/(1-\psiii)$ to be recovered relatively well even with power-law models. Let us verify this prediction by plotting the inferred value of $\psiiii/(1-\psiii)$ versus the true values, in \Fref{fig:psicomb}.
As expected, there is a good match between the true and recovered values.
The few outliers with respect to the equality line are lenses with a very asymmetric image configuration, for which \Eref{eq:radmagrat} is no longer accurate.

\subsection{Inferences on $H_0$}

Let us now turn the attention to the inference on the Hubble constant.
For each lens, given the model parameters obtained from the fit to the lensing data, we can calculate the model time delay as a function of $H_0$, $\Delta t(H_0)$, using \Eref{eq:dt}:
\begin{multline}
\Delta t (H_0, \gamma, \tein, \beta_s) = \frac{\tilde{H_0}}{H_0}\frac{D_{\Delta t}(\tilde{H_0})}{c} \times \\
\left[\frac{(\theta_B - \beta_s)^2}{2} - \psi(\theta_B) - \frac{(\theta_A - \beta_s)^2}{2} + \psi(\theta_A)\right],
\end{multline}
where $\tilde{H_0}$ is a fiducial value of the Hubble constant, $D_{\Delta t}(\tilde{H_0})$ the time delay distance calculated for $H_0=\tilde{H_0}$, and the image positions and lens potential are a function of the lens model parameters $\gamma$, $\tein$ and $\beta_s$.

The inferred value of $H_0$ is then given by the following:
\begin{equation}
H_0 = \tilde{H_0} \frac{\Delta t (\tilde{H_0}, \gamma, \tein, \beta_s)}{\Delta t^{\mathrm{(obs)}}},
\end{equation}
where $\Delta t^{\mathrm{(obs)}}$ is the observed value of the time delay, which is assumed to be known exactly.

\begin{figure}
 \includegraphics[width=\columnwidth]{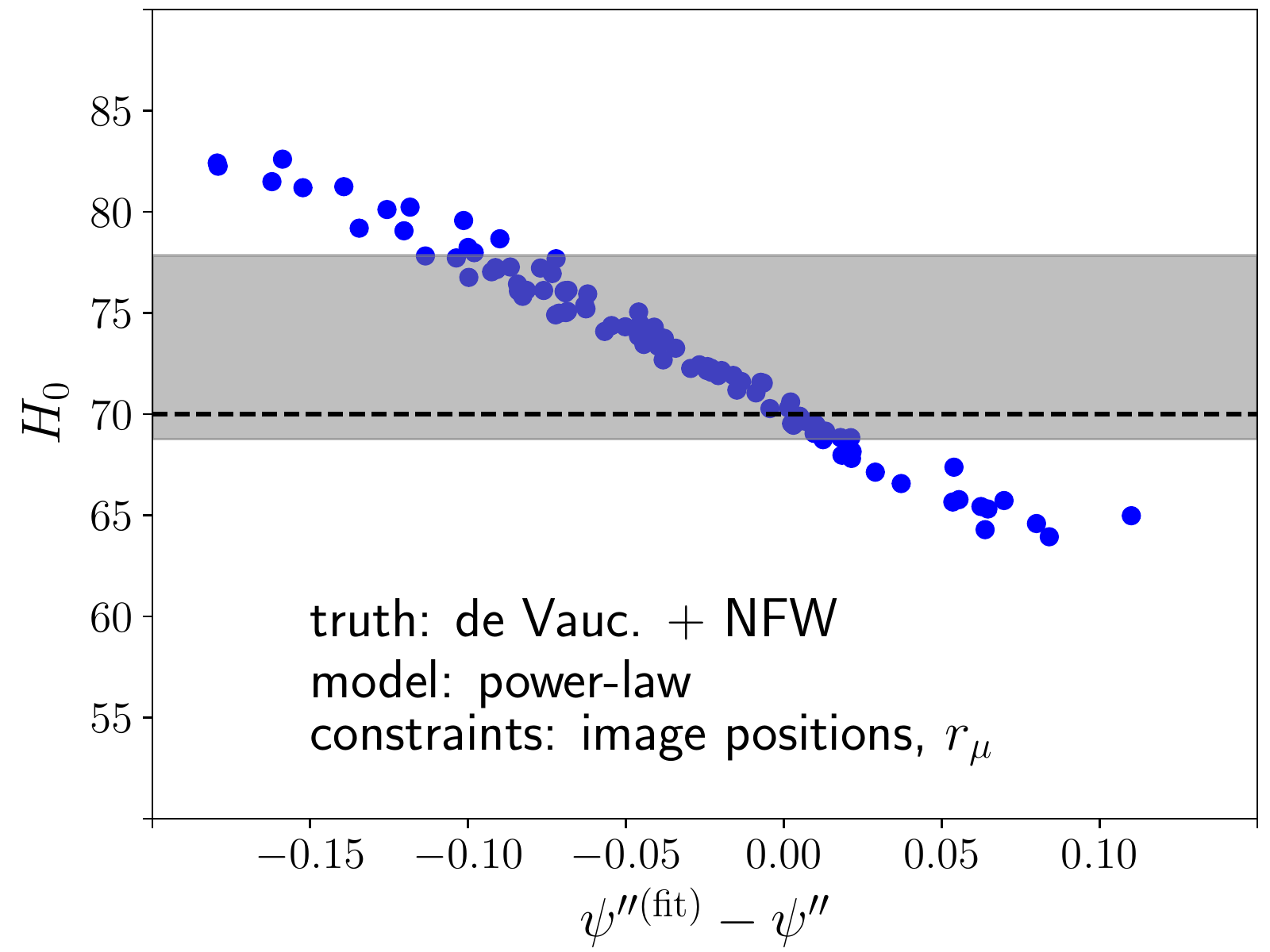}
 \caption{Inference on $H_0$ for each lens, obtained by fitting a power-law model to image position and radial magnification data, and to time delay measurements with no uncertainty.
The quantity on the $x$ axis is the difference between the value of $\psi''$ obtained from the power-law fit to the lensing data and the true value of $\psi''$.
The dashed horizontal line marks the true value of $H_0$ used to create the mock ($70\,\rm{km}\,\rm{s}^{-1}\,\rm{Mpc}^{-1}$).
The grey horizontal band shows the region containing 68\% of the $H_0$ measurements.
}
 \label{fig:plH0}
\end{figure}

The resulting inference on $H_0$ obtained from each lens is plotted in \Fref{fig:plH0}.
The recovered values of $H_0$ are plotted as a function of the difference between the value of $\psiii$ inferred from the power-law fit to the data and the true value of $\psiii$.
The bias on $H_0$ should correlate with this quantity: since the time delay is, to lowest order in image displacement from the Einstein radius, sensitive to the second derivative of the potential, lenses for which the power-law assumption leads to a wrong inference of $\psiii$ will be the ones with the largest bias.
As expected, we can see a clear trend between $\psi''^{\mathrm{(fit)}} - \psiii$ and the inferred value of $H_0$, though with some scatter.
The average value of the inferred Hubble constant is $73.3\Hunit$, with a $4.6\Hunit$ standard deviation: the power-law assumption introduces an average bias of $\sim5\%$ on $H_0$, but the bias is significantly larger for an appreciable fraction of the objects.

The average and scatter of the bias are somewhat arbitrary, as they depend on the details of the mock lens population used for the test. 
For instance, \citet{Xu++16} found a significantly larger scatter in the bias on $H_0$, with a similar experiment based on mock lenses extracted from cosmological hydrodynamical simulations.
A significant fraction of objects in the \citet{Xu++16} sample are lenses with a power-law density slope much flatter than isothermal ($\gamma \ll 2$, corresponding to $s_\lambda \ll 1$ in their notation). Those objects corresponds to very large biases on $H_0$.
The mock sample introduced in this Section, instead, has a narrower distribution in power-law slope $\gamma$, as can be seen from the y-axis in \Fref{fig:slope}: the average value of $\gamma$ is $1.92$, with a $0.15$ scatter. This is closer to the typical values of the slope measured for time delay lenses, compared to the \citet{Xu++16} sample \citep[see e.g.][]{Suy++10, Suy++13, Agn++16, Won++17}.
This test shows that, as argued by \citet{Xu++16}, even with lens profiles inferred to be close to isothermal, the bias on $H_0$ can be significant.

\section{Flexible models}\label{sect:gnfw}

Power-law models, fitted to image position and radial magnification ratios, cannot provide a very accurate estimate of the time-delay of the lenses in the mock sample. This is because the value of the second derivative of the lens potential, crucial for the determination of the time delay, is not directly constrained by the data, but rather imposed by a fixed relation with the first and third derivative (the curve in \Fref{fig:psi}), which are the quantities lensing observables are most sensitive to. 
In order to gain in accuracy, a more flexible model family, able to describe a range of values of $\psiii$ given $\psiiii$, and thus covering entirely the region of \Fref{fig:psi} where mock lenses lie, is needed.
In other words, it is necessary to add at least one degree of freedom in the radial density profile.

Let us consider a composite model, with a de Vaucouleurs profile describing the stellar component and a generalised NFW profile (gNFW) to describe the dark matter halo \citep{Zha96}.
\begin{equation}
\rho_{\mathrm{gNFW}}(r) = \frac{\rho_0}{(r/r_s)^{\gammadm} (1 + r/r_s)^{3 - \gammadm}}.
\end{equation}
The density profile of a gNFW profile has an inner slope of $\gammadm$ and falls off as $r^{-3}$ at large radii. It reduces to a pure NFW profile for $\gammadm=1$.

I assume that the effective radius of the de Vaucouleurs profile is known exactly.
This assumption leaves one degree of freedom for the stellar component: its total mass.
The dark matter component has three free parameters: mass, inner slope $\gammadm$, and scale radius $r_s$.
I wish to work with a family of models with only three degrees of freedom in total.
Therefore, I arbitrarily fix the scale radius of the halo to ten times the effective radius:
\begin{equation}\label{eq:scaleradius}
r_s = 10 \reff.
\end{equation}
Due to this choice, this model cannot provide a perfect description of the galaxies in the mock: although mock halos have an NFW density profile, their scale radius is assigned through a mass-concentration relation with scatter. Therefore they do not satisfy \Eref{eq:scaleradius}.
Nevertheless, as I will show, this model can still allow us to make an accurate prediction on the time delay, thanks to its flexibility.

I have increased the number of degrees of freedom in the radial density profile from two to three.
Obviously, this new model cannot be fully constrained with only two image positions and one measurement of the radial magnification ratio.
In order to make an inference on the time delay, additional constraints are needed.
I consider the central velocity dispersion as a new observable.
Velocity dispersion information is often used in time delay cosmology measurements to improve the precision and accuracy of the model \citep{Suy++14,Won++17}.

For each mock lens, I use the spherical Jeans equation to calculate its surface brightness-weighted line-of-sight velocity dispersion, integrated within a circular aperture of radius $\reff/2$, which we denote $\sigmae$. I assume isotropic orbits. I do not convolve the velocity dispersion by an artificial point spread function, for the sake of computational time.
This observable is also assumed to be known exactly. Details relative to the velocity dispersion model are given in Appendix~\ref{sect:appendixd}.

I fit the composite model to image position, radial magnification ratio and velocity dispersion of each mock lens. For a given set of parameter values, I calculate a model velocity dispersion in the same way the data was generated, by using the spherical Jeans equation under the assumption of isotropic orbits. Essentially, I am assuming that the line-of-sight structure and orbital anisotropy profile of the lens is known exactly, which is a very optimistic assumption.
With four constraints (two image positions, one radial magnification ratio and one velocity dispersion) and four parameters (source position, stellar mass, dark matter mass and inner dark matter slope), an exact solution to the problem can be found.

\begin{figure}
 \includegraphics[width=\columnwidth]{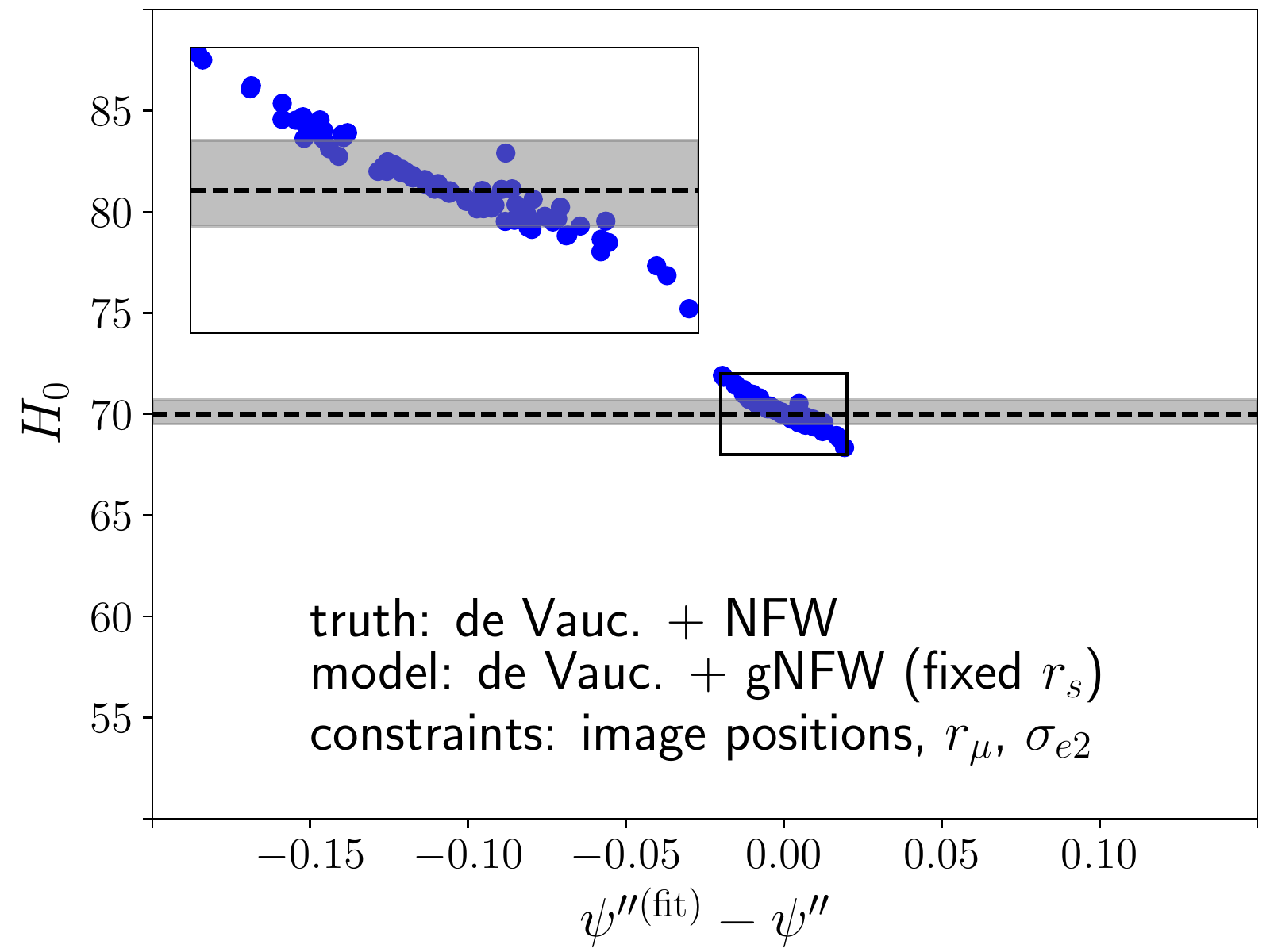}
 \caption{Inference on $H_0$ for each lens, obtained by fitting a composite de Vaucouleurs + gNFW model to image position, radial magnification ratio, velocity dispersion and time delay measurements.
The quantity on the $x$ axis is the difference between the inferred value of $\psiii$ and the true value. The dashed horizontal line marks the true value of $H_0$ used to create the mock ($70\,\rm{km}\,\rm{s}^{-1}\,\rm{Mpc}^{-1}$).
The grey horizontal band shows the mean inference on $H_0$, with the region enclosing 68\% of the measurements.
The scale of the plot is the same as that of \Fref{fig:plH0}. The inset in the upper left shows in greater detail the region enclosed by the central rectangle.
}
 \label{fig:gnfw_indH0}
\end{figure}

I fit all 100 lenses in the mock, then, similarly to \Sref{sect:mock}, I obtain an inference on $H_0$ for each lens.
Individual $H_0$ measurements and the region enclosing 68\% of the data points are plotted in \Fref{fig:gnfw_indH0}.
The average bias on the Hubble constant is $0.1\Hunit$, and the scatter is as low as $0.6\Hunit$: $H_0$ is recovered at better than 1\% accuracy.
At the same time, the values of the second derivative of the potential are recovered with high accuracy (data points are clustered around zero on the x-axis of \Fref{fig:gnfw_indH0}), compared to the power-law case.
This is a nontrivial result, because 1) none of the galaxies in the mock can be described exactly by the model used to perform the fit, and 2) the velocity dispersion information is, in general, probing the gravitational potential of the lens galaxy in a different region with respect to the scale relevant for the estimate of the time delay (i.e. the half-light radius vs. the Einstein radius).

The excellent agreement between the value of $H_0$ obtained with the flexible model thanks to the addition of velocity dispersion information raises a question: would it be possible to make a similarly accurate inference using power-law profiles, if we replace the radial magnification constraint with the velocity dispersion?
As pointed out earlier, the radial magnification ratio is mostly sensitive to the third derivative of the potential. Although this piece of information can be used to get a high precision measurement of the power-law slope $\gamma$, it does not help in the prediction of the time delay, since an accurate knowledge of $\psiii$ is required to constrain the latter.

I then revert to power-law models and fit them to image position and velocity dispersion data.
Since two image positions and a velocity dispersion are sufficient to constrain a power-law model, I will ignore radial magnification information for this test.
In real observations, all available data is fitted at the same time, and the best fit model is the result of a combination of lensing and stellar kinematics constraints, with different relative weights depending on the uncertainties of the different measurements \citep[see e.g.][for an example of how velocity dispersion changes the inference with respect to a lensing-only fit]{Suy++14}.
This test corresponds to a limiting case in which the uncertainties on the velocity dispersion and image position measurements are much smaller than those on the surface brightness distribution of the quasar host galaxy, while the opposite limit corresponds to the test of \Sref{sect:mock}.

\begin{figure}
 \includegraphics[width=\columnwidth]{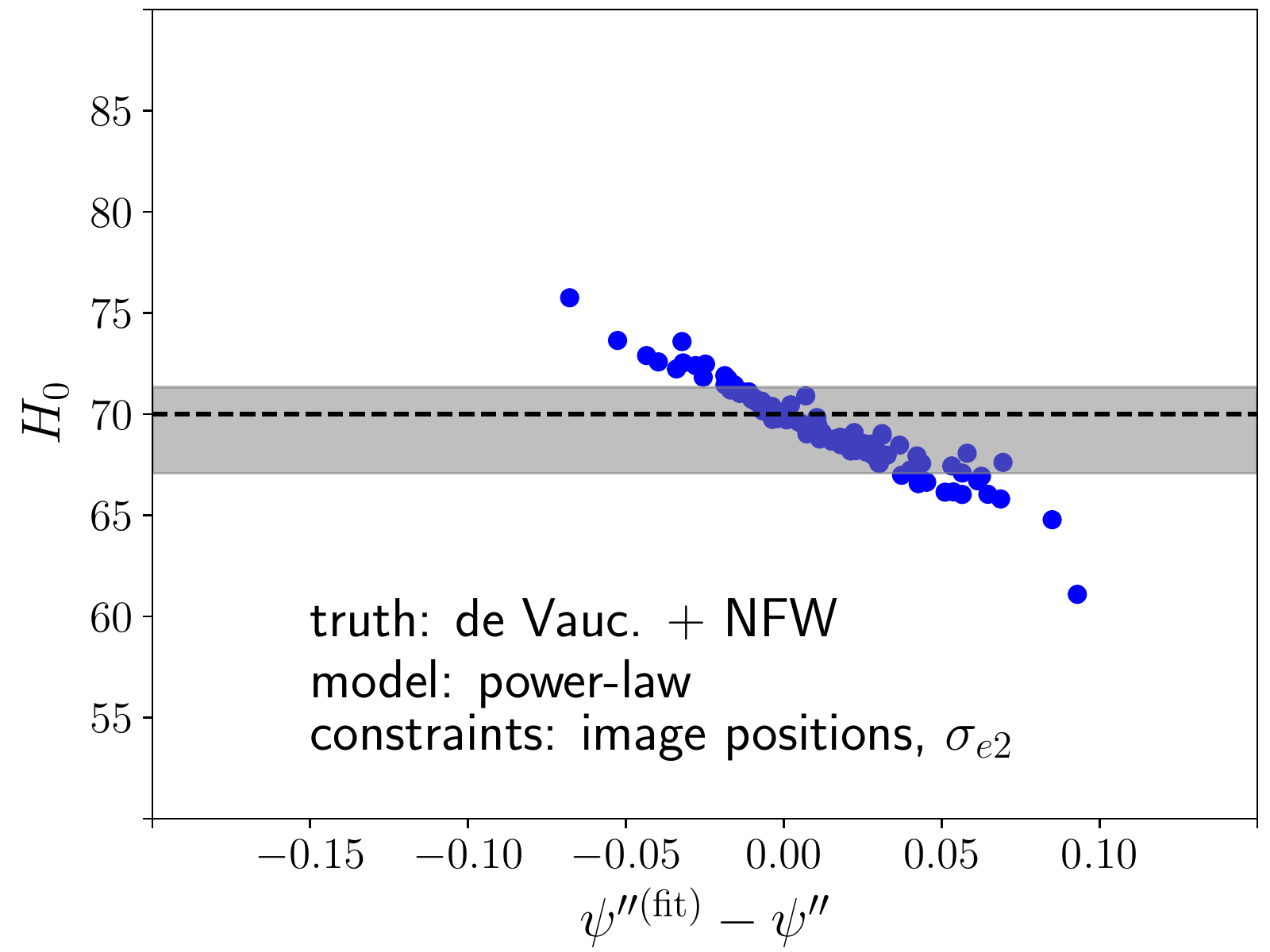}
 \caption{Inference on $H_0$ for each lens, obtained by fitting a power-law model to image position, velocity dispersion, and to time delay measurements.
The quantity on the $x$ axis is the difference between the inferred value of $\psiii$ and the true value. The dashed horizontal line marks the true value of $H_0$ used to create the mock ($70\,\rm{km}\,\rm{s}^{-1}\,\rm{Mpc}^{-1}$).
The grey horizontal band shows the weighted mean inference on $H_0$, with its 68\% uncertainty region.
The scale of the plot is the same as that of \Fref{fig:plH0}.}
 \label{fig:pl_indH0_wdyn}
\end{figure}

The individual inferences on $H_0$ and the region enclosing 68\% of the measurements are plotted in \Fref{fig:pl_indH0_wdyn}.
The average value of the Hubble constant is now $69.2\Hunit$, with a $2.1\Hunit$ standard deviation, a significant improvement compared to the values obtained in \Sref{sect:pl} with purely lensing constraints.
A comparison between \Fref{fig:plH0} and \Fref{fig:pl_indH0_wdyn} suggests that stellar kinematics provides a more accurate estimate of the second derivative of the lens potential, compared to radial magnification information, since the spread of the points along the x-axis is significantly reduced in this new test.
The more accurate inference on $H_0$ is then a direct consequence of the improvement in the accuracy of the measurement of $\psiii$.
\section{Discussion}\label{sect:discuss}

On the basis of simple analytical arguments, I predicted in \Sref{sect:pot} that, in order to model time delays with an accuracy of a percent, a class of lens models with at least three degrees of freedom in the radial profile is needed.
This conclusion cast doubts on the adequacy of power-law models in time delay cosmography studies.
Tests performed on mock lensing observations showed that using power-law models can lead to biases as high as $\sim10\%$ on the inference of the Hubble constant (see \Fref{fig:plH0}).
However, 
fitting the same power-law models to the quasar image position and velocity dispersion data, and ignoring magnification information, gives a much more accurate answer.

The problem does not lie with power-law models themselves, but rather with the data used to constrain their parameters.
In detailed lensing studies, the slope of the lens density profile is typically measured by fitting a power-law model to the surface brightness distribution of the arcs.
However, as argued in \Sref{sect:pot}, this is only a measurement of radial magnification ratio, which to first nontrivial order is mostly sensitive to the third derivative of the potential.
The second derivative of the potential, more important for the determination of the time delay, is then assigned by a fixed relation, \Eref{eq:psiiipl_given_psiiii}, introducing a bias.

A possible alternative, or complementary information, to radial magnification ratios, is the use of stellar kinematics.
\cite{STA17} recently showed how stellar kinematics can be used to improve the precision in the determination of cosmological parameters.
In this study I use velocity dispersion and relax the assumption of a power-law density profile to improve the accuracy of the inference.

The tests performed in \Sref{sect:gnfw} are promising: by fitting a composite model with three degrees of freedom to lensing and velocity dispersion constraints, $H_0$ can be recovered with 1\% accuracy.
In particular, the results show that it is not necessary to recover the dark matter distribution exactly in order to obtain an accurate measurement of $H_0$: the model for the dark matter halo consists of a gNFW profile with $r_s=10\reff$, while the true dark matter profile of the mock is an NFW with a variable $r_s/\reff$ ratio.
However, this result was obtained under the assumption of a perfect knowledge of the stellar density profile, with the only unknown parameter of the stellar component being the mass-to-light ratio.
In real situations, this is difficult to achieve, due to the presence of radial gradients in the stellar population of galaxies.

Additionally, I assumed spherical symmetry and isotropic orbits.
In real applications, projection effects and orbital anisotropy can introduce large biases if not carefully accounted for. For instance, \citet{Bir++16} found a 15\% change in the inferred value of $H_0$ just by changing the prior on the anisotropy, in their analysis of the lens RXJ1131$-$1231. 
Spatially resolved kinematic information, such as velocity measurements from integral field spectroscopy, is most likely necessary for an accuracy goal of a few percent.
I refer to \citet{STA17} for a more realistic study of how integral field spectroscopy observations can be used in combination with lensing to constrain lens models for time delay cosmology.

All of the tests carried out in this study are based on a strong assumption: that of circular symmetry.
Although the main analytical results derived in \Sref{sect:pot} also hold in cases of small departures from axisymmetry,
an important consequence of the circular symmetry assumption is the absence of quadruply imaged (quad) systems from the mock sample.
Quad lenses provide more constraints compared to doubles, with two more image positions and two more time delays.
Indeed, \citet{Suy++14} showed that some model degeneracies can be broken by simultaneously fitting three time delays between the four images. 
Therefore, the results of the present analysis are not directly applicable to quads.

\section{Conclusions}\label{sect:concl}

I revisited the problem of the mass-sheet degeneracy in time delay lensing studies.
By means of a Taylor expansion around the Einstein radius, I studied how different lensing observables depend on the local derivatives of the lens potential, in the case of circular lenses.
An important finding is that image position and radial magnification ratio, typically used to constrain lens mass models, are only sensitive to a fixed combination of the third and second derivative of the potential, $\psiiii/(1-\psiii)$.
On the other hand, time delay depends on the first and second derivative of the lens potential, to lowest order in the distance from the critical curve.

Using image positions and radial magnification ratios as constraints on a lens model results in a precise inference on the quantity $\psiiii/(1-\psiii)$.
However, the assumption of a power-law density profile imposes a fixed relation between $\psiii$ and $\psiiii$, which can introduce a bias on the predicted time delay.

I tested the accuracy of the inference on $H_0$ with power-law lens models by fitting mock observations, finding an average bias of 5\%, with 6\% scatter.
I then showed that by adding one degree of freedom to the radial density profile, and by adding velocity dispersion constraints, $H_0$ can be recovered with a 1\% accuracy.
Interestingly, power-law models can still provide an $H_0$ estimate with an accuracy within 3\% if velocity dispersion information is used to constrain the slope, instead of radial magnification ratios, although my treatment of the velocity dispersion modelling is based on very optimistic assumptions.

I conclude that accurate stellar kinematics information is key in order to make correct inferences with time delay lensing. Given the current precision of time delay measurements, power-law profiles are still an adequate model for studies of double lenses, provided that stellar kinematics information is used to constrain their density slope.
In order to reach an accuracy goal of 1\% on $H_0$, I recommend 1) the use of lens models with at least three degrees of freedom in the radial density profile, and 
2) controlled experiments in lens modelling, to robustly determine which properties of a lens can be reliably constrained by fitting the surface brightness distribution of a lensed extended source.

\section*{acknowledgments}
I wish to thank Phil Marshall and Adriano Agnello for useful discussions, and the anonymous referee for valuable suggestions that helped improve the quality of the paper.
This work was supported by World Premier International Research Center Initiative (WPI Initiative), MEXT, Japan.
AS is partly supported by KAKENHI Grant Number JP17K14250. 
\bibliographystyle{mnras}
\bibliography{references}

\appendix
\onecolumn
\section{Taylor expansion of the image positions}\label{sect:appendixa}
\Eref{eq:lens2nd} has been obtained by a Taylor expansion of \Eref{eq:quadratic} up to second order in $\Delta\theta_A$. Here I show its derivation. Let us focus on the square root term in \Eref{eq:quadratic}:
\begin{equation}\label{eq:sqrt}
s(\Delta\theta_A) = \sqrt{1 + 2\frac{\psiiii}{1-\psiii}\Delta\theta_A - \left(\frac{\psiiii}{1-\psiii}\right)^2\Delta\theta_A^2}.
\end{equation}
Its Taylor series up to order $\Delta\theta_A^2$ is given by
\begin{equation}\label{eq:sqrttaylor}
s(\Delta\theta_A) = 1 + \left. \frac{\partial s}{\partial \Delta\theta_A} \right\rvert_{\Delta\theta_A=0}\Delta\theta_A + \frac12\left.\frac{\partial^2 s}{\partial \Delta\theta_A^2}\right\rvert_{\Delta\theta_A=0}\Delta\theta_A^2.
\end{equation}
The first derivative of \Eref{eq:sqrt} is given by
\begin{equation}\label{eq:firstderiv}
\frac{\partial s}{\partial \Delta\theta_A} = \frac12\left[1 + 2\frac{\psiiii}{1-\psiii}\Delta\theta_A - \left(\frac{\psiiii}{1-\psiii}\right)^2\Delta\theta_A^2\right]^{-1/2}\left[2\frac{\psiiii}{1-\psiii}- 2\left(\frac{\psiiii}{1-\psiii}\right)^2\Delta\theta_A\right].
\end{equation}
The second derivative is
\begin{align}\label{eq:secondderiv}
\frac{\partial^2 s}{\partial \Delta\theta_A^2} = & -\frac14\left[1 + 2\frac{\psiiii}{1-\psiii}\Delta\theta_A - \left(\frac{\psiiii}{1-\psiii}\right)^2\Delta\theta_A^2\right]^{-3/2}\left[2\frac{\psiiii}{1-\psiii}- 2\left(\frac{\psiiii}{1-\psiii}\right)^2\Delta\theta_A\right]^2 + \nonumber \\ 
& - \left[1 + 2\frac{\psiiii}{1-\psiii}\Delta\theta_A - \left(\frac{\psiiii}{1-\psiii}\right)^2\Delta\theta_A^2\right]^{-1/2}\left(\frac{\psiiii}{1-\psiii}\right)^2.
\end{align}
Evaluating \Eref{eq:firstderiv} and \Eref{eq:secondderiv} at $\Delta\theta_A=0$ and substituting into \Eref{eq:sqrttaylor} gives
\begin{equation}\label{eq:sqrttaylor2nd}
s(\Delta\theta_A) = 1 + \frac{\psiiii}{1-\psiii}\Delta\theta_A - \left(\frac{\psiiii}{1-\psiii}\right)^2\Delta\theta_A^2.
\end{equation}
Substituting \Eref{eq:sqrttaylor2nd} into \Eref{eq:quadratic}, and keeping only the positive solution produces \Eref{eq:lens2nd}.

\section{Taylor expansion of the radial magnification ratio}\label{sect:appendixb}
I derive here the 2nd order Taylor expansion of the radial magnification ratio leading to \Eref{eq:radmagrat}.
The ratio of radial magnifications between image A and B is
\begin{equation}\label{eq:numden}
\frac{\mu_{r,A}}{\mu_{r,B}} = \dfrac{1 - \dfrac{d\alpha(\theta_B)}{d\theta}}{1 - \dfrac{d\alpha(\theta_A)}{d\theta}}.
\end{equation}
Let us consider the denominator of the above equation: the radial magnification at image A. By combining \Eref{eq:murdef} and \Eref{eq:dalphaa}, this is, to second order in $\Delta\theta_A$,
\begin{equation}\label{eq:murtaylor1}
\mu_{r,A} = \left(1 - \psiii - \psiiii\Delta\theta_A - \psiiv\Delta\theta_A^2\right)^{-1}.
\end{equation}
Its Taylor series up to order $\Delta\theta_A^2$ is defined as
\begin{equation}\label{eq:murtaylor2}
\mu_{r,A} = \mu_{r,A}(\tein) + \left. \frac{\partial \mu_{r,A}}{\partial \Delta\theta_A} \right\rvert_{\Delta\theta_A=0}\Delta\theta_A + \frac12\left.\frac{\partial^2 \mu_{r,A}}{\partial \Delta\theta_A^2}\right\rvert_{\Delta\theta_A=0}\Delta\theta_A^2.
\end{equation}
From \Eref{eq:murtaylor1}, the first derivative of the radial magnification is
\begin{equation}\label{eq:dmura}
\frac{\partial \mu_{r,A}}{\partial \Delta\theta_A} = -\left(1 - \psiii - \psiiii\Delta\theta_A - \psiiv\Delta\theta_A^2\right)^{-2}(-\psiiii - 2\psiiv\Delta\theta_A).
\end{equation}
The second derivative is
\begin{equation}\label{eq:d2mura}
\frac{\partial^2 \mu_{r,A}}{\partial \Delta\theta_A^2} = 2\left(1 - \psiii - \psiiii\Delta\theta_A - \psiiv\Delta\theta_A^2\right)^{-3}(-\psiiii - 2\psiiv\Delta\theta_A)^2 + 2\left(1 - \psiii - \psiiii\Delta\theta_A - \psiiv\Delta\theta_A^2\right)^{-2}\psiiv.
\end{equation}
Evaluating \Eref{eq:murtaylor1}, \Eref{eq:dmura} and \Eref{eq:d2mura} at $\Delta\theta_A=0$ and substituting into \Eref{eq:murtaylor2} gives
\begin{equation}\label{eq:mura}
\mu_{r,A} = \frac{1}{1-\psiii}\left[1 + \frac{\psiiii}{1-\psiii}\Delta\theta_A + \left(\frac{\psiiii}{1-\psiii}\right)^2\Delta\theta_A^2 + \frac{\psiiv}{1-\psiii}\Delta\theta_A^2\right].
\end{equation}
Let us now focus on the numerator of \Eref{eq:numden}, the radial magnification of image B.
Making use of \Eref{eq:dalphab}, this is
\begin{equation}
\frac{1}{\mu_{r,B}} = 1 - \psiii + \psiiii\Delta\theta_B + \psiiv\Delta\theta_B^2.
\end{equation}
Let us express $\Delta\theta_B$ in terms of $\Delta\theta_A$, using \Eref{eq:lens2nd}, keeping terms up to $\Delta\theta_A^2$:
\begin{equation}\label{eq:eralora}
\frac{1}{\mu_{r,B}} = 1 - \psiii + \psiiii\Delta\theta_A - \frac{\psi'''^2}{1-\psiii}\Delta\theta_A^2 - \psiiv\Delta\theta_A^2.
\end{equation}
Finally, plugging \Eref{eq:mura} and \Eref{eq:eralora} into \Eref{eq:numden} gives
\begin{equation}
\frac{\mu_{r,A}}{\mu_{r,B}} = \left[1 + \frac{\psiiii}{1-\psiii}\Delta\theta_A - \left(\frac{\psiiii}{1-\psiii}\right)^2\Delta\theta_A^2 - \frac{\psiiv}{1-\psiii}\Delta\theta_A^2\right]\left[1 + \frac{\psiiii}{1-\psiii}\Delta\theta_A + \left(\frac{\psiiii}{1-\psiii}\right)^2\Delta\theta_A^2 + \frac{\psiiv}{1-\psiii}\Delta\theta_A^2\right].
\end{equation}
Expanding the product and keeping only terms up to $\Delta\theta_A^2$ gives \Eref{eq:radmagrat}.

\section{Taylor expansion of the time delay}\label{sect:appendixc}
I will illustrate the steps leading to the Taylor expansion of the time delay to second order in $\Delta\theta_A$, \Eref{eq:dt2nd}.
The time delay between image A and B is given by \Eref{eq:dt}. Noting that the quantities $(\theta_A- \beta_s)$ and $(\theta_B - \beta_s)$ are equal to the deflection angles in A and B (see \Eref{eq:lensequation}), this can be rewritten as
\begin{equation}\label{eq:dttaylor}
\Delta t = \frac{D_{\Delta t}}{c}\left[\frac{\alpha_B^2}{2} - \psi(\theta_B) - \frac{\alpha_A^2}{2} + \psi(\theta_A)\right].
\end{equation}
Given \Eref{eq:potA}, the Taylor series of the deflection angle at image A, up to order $\Delta\theta_A^2$, is
\begin{equation}
\alpha_A = \psii + \psiii\Delta\theta_A + \frac12\psiiii\Delta\theta_A^2.
\end{equation}
Its square is, up to order $\Delta\theta_A^2$,
\begin{equation}
\alpha_A^2 = \psi'^{2} + 2\psii\psiii\Delta\theta_A + \psiii^{2}\Delta\theta_A^2 + \psii\psiiii\Delta\theta_A^2.
\end{equation}
Similarly, for image B,
\begin{equation}\label{eq:alphaB}
\alpha_B = -\psii + \psiii\Delta\theta_B - \frac12\psiiii\Delta\theta_B^2,
\end{equation}
where I have used \Eref{eq:potB}. Let us express \Eref{eq:alphaB} in terms of $\Delta\theta_A$, using \Eref{eq:lens2nd} and discarding terms higher than 2nd order:
\begin{equation}\label{eq:alphaBbythetaA}
\alpha_B = -\psii + \psiii\Delta\theta_A - \frac{\psiii\psiiii}{1-\psiii}\Delta\theta_A^2 - \frac12\psiiii\Delta\theta_A^2.
\end{equation}
Let us take the square and eliminating all terms of order higher than $\Delta\theta_A^2$:
\begin{equation}
\alpha_B^2 = \psi'^{2} - 2\psii\psiii\Delta\theta_A + \psi''^{2}\Delta\theta_A^2 + 2\frac{\psii\psiii\psiiii}{1-\psiii}\Delta\theta_A^2 + \psii\psiiii\Delta\theta_A^2.
\end{equation}
The difference between the squares of the two deflection angles, entering \Eref{eq:dttaylor}, is 
\begin{equation}\label{eq:pacco1}
\alpha_B^2 - \alpha_A^2 = -4\psii\psiii\Delta\theta_A + 2\frac{\psii\psiii\psiiii}{1-\psiii}\Delta\theta_A^2.
\end{equation}
Let us now focus on the potential terms in \Eref{eq:dttaylor}.
Let us express the potential at image B in terms of $\Delta\theta_A$, using \Eref{eq:potB} and \Eref{eq:lens2nd}, and keeping terms up to $\Delta\theta_A^2$:
\begin{equation}
\psi(\theta_B) = \psi(\tein) - \psii\Delta\theta_A + \frac{\psii\psiiii}{1-\psiii}\Delta\theta_A^2 + \frac12\psiii\Delta\theta_A^2.
\end{equation}
The difference between the potential at image A and image B is
\begin{equation}\label{eq:pacco2}
\psi(\theta_A) - \psi(\theta_B) = 2\psii\Delta\theta_A - \frac{\psii\psiiii}{1-\psiii}\Delta\theta_A^2.
\end{equation}
Substituting \Eref{eq:pacco1} and \Eref{eq:pacco2} into \Eref{eq:dttaylor} gives \Eref{eq:dt2nd}.

\section{Tangential critical curve for an elliptical potential}\label{sect:appendixe}
The Jacobian matrix of the mapping between lens and source plane, given a lens potential $\Phi$, is
\begin{equation}
A = \frac{\partial\boldsymbol\beta}{\partial\boldsymbol\theta} = \left(\begin{matrix} 1 - \dfrac{\partial^2\Phi}{\partial \theta_1^2} & \dfrac{\partial^2\Phi}{\partial\theta_1\partial\theta_2} \\ \dfrac{\partial^2\Phi}{\partial\theta_1\partial\theta_2} & 1 - \dfrac{\partial^2\Phi}{\partial \theta_2^2} \end{matrix}\right).
\end{equation}
For a potential with elliptical symmetry, as specified by \Eref{eq:potell}, its derivatives are
\begin{align}
\frac{\partial^2\Phi}{\partial\theta_1^2} = & \left[\frac{q}{\rho} - \frac{q^2\theta_1^2}{\rho^3}\right]\frac{\partial\phi}{\partial\rho} + \frac{q^2\theta_1^2}{\rho^2}\frac{\partial^2\phi}{\partial\rho^2}, \\
\frac{\partial^2\Phi}{\partial\theta_2^2} = & \left[\frac{1}{q\rho} - \frac{\theta_2^2}{q^2\rho^3}\right]\frac{\partial\phi}{\partial\rho} + \frac{\theta_2^2}{q^2\rho^2}\frac{\partial^2\phi}{\partial\rho^2}, \\
\frac{\partial^2\Phi}{\partial\theta_1\partial\theta_2} = & \frac{\theta_1\theta_2}{\rho^2}\left[\frac{\partial^2\phi}{\partial\rho^2} - \frac{1}{\rho}\frac{\partial\phi}{\partial\rho}\right].
\end{align}
Critical curves correspond to solutions of the equation
\begin{equation}
\det{A} = 0.
\end{equation}
I wish to determine where the tangential critical curve crosses the $\theta_1$ axis.
For $\theta_2=0$, the cross terms of the Jacobian are zero, and the eigenvalues correspond to the diagonal terms. The tangential eigenvalue, in particular, is the term $A_{22}$, for $\theta_2=0$. The tangential critical curve then corresponds to
\begin{equation}
\left[\frac{1}{q\rho(\theta_1, 0)}\right]\frac{\partial\phi}{\partial\rho} = 1,
\end{equation}
which is equivalent to \Eref{eq:critell}.

\section{Velocity dispersion}\label{sect:appendixd}
I will describe how the model velocity dispersions used in the tests of \Sref{sect:gnfw} are computed.
Given a galaxy with 3D enclosed mass profile $M(r)$ and a 3D density of tracers (stars) $\rho_*(r)$, I use the spherical Jeans equation to define its stellar velocity dispersion:
\begin{equation}
\frac{1}{\rho_*}\frac{d\rho_*\sigma_r^2}{dr} + 2\frac{\sigma_\theta^2}{r} = -\frac{GM(r)}{r^2},
\end{equation}
where $\sigma_r$ and $\sigma_\theta$ are the radial and tangential components of the velocity dispersion tensor \citep{B+T87}.
I then assume isotropic orbits,
\begin{equation}
\sigma_\theta = \sigma_r,
\end{equation}
to obtain
\begin{equation}
\frac{\rho_*(r)\sigma_r^2(r)}{G} = \int_r^\infty \frac{\rho_*(s) M(s)}{s^2}ds.
\end{equation}
The line-of-sight velocity dispersion at a projected distance $R$ from the galaxy centre is given by
\begin{equation}\label{eq:sigmalos}
I(R)\sigma_{\mathrm{los}}^2(R) = 2G \int_R^\infty \frac{\sqrt{r^2 - R^2}}{r^2}\rho_*(r)M(r)dr,
\end{equation}
where $I(R)$ is the surface brightness \citep{P+S97}. The surface brightness-weighted velocity dispersion within $\reff/2$ is calculated by integrating \Eref{eq:sigmalos} over an aperture of radius $\reff/2$, and dividing it by the integrated brightness within the same aperture.

\end{document}